matrix representation of% ****** Start of file apssamp.tex ******
\documentclass[%
superscriptaddress,
showkeys,
nofootinbib,
 amsmath,amssymb,
 aps,
]{revtex4-2}

\usepackage{graphicx}% Include figure files
\usepackage{dcolumn}% Align table columns on decimal point
\usepackage{bm}% bold math
\usepackage{amsthm}
\usepackage{floatrow}

\usepackage{color}
\usepackage[T1]{fontenc}
\newtheorem{lemma}{Lemma}
\newtheorem{definition}{Definition}
\newtheorem{theorem}{Theorem}
\newtheorem{proposition}{Proposition}

\newcommand{\bra}[1]{\langle {#1} |}
\newcommand{\ket}[1]{| {#1} \rangle}
\newcommand{\braket}[2]{\langle {#1} |{#2} \rangle}
\newcommand{\ketbra}[1]{| {#1} \rangle\langle {#1} |}

\newcommand{\lpnorm}[2]{\left\|#2\right\|_{#1}}%Lp norm
\newcommand{\trdist}[1]{\left\|#1\right\|_{\text{tr}}}%Tr distance
\newcommand{\fidelity}[2]{F\left(#1,#2\right)}
\newcommand{\tr}[1]{tr\left[#1\right]}%trace
\newcommand{\ptr}[2]{\text{tr}_{#1}\left[#2\right]}%partial trace
\newcommand{\conv}[1]{\text{conv}\left(#1\right)}%trace

\newcommand{\linop}[1]{\mathbf{L}\left(#1\right)}
\newcommand{\pos}[1]{\mathbf{Pos}\left(#1\right)}
\newcommand{\unitary}[1]{\mathbf{U}\left(#1\right)}
\newcommand{\dop}[1]{\mathbf{S}\left(#1\right)}%set of density operator
\newcommand{\puredop}[1]{\mathbf{P}\left(#1\right)}%set of rank-1 density operator

\newcommand{\hh}{\mathcal{H}}%Hilbert space
\newcommand{\vv}{\mathcal{V}}%subspace of Hilbert space
\newcommand{\cd}{\mathbb{C}^d}%Hilbert space
\newcommand{\cdim}[1]{\mathbb{C}^{#1}}
\newcommand{\rdim}[1]{\mathbb{R}^{#1}}
\newcommand{\vspan}[1]{\text{span}\left(#1\right)}
\newcommand{\nn}{\mathbb{N}}%natural number}
\newcommand{\rr}{\mathbb{R}}%real number}
\newcommand{\cc}{\mathbb{C}}%complex number}

\newcommand{\re}[1]{\text{Re}\left(#1\right)}%real part

\newcommand{\phiI}[1]{\phi_{#1}}
\newcommand{\psiI}[1]{\psi_{#1}}

\newcommand{\indicator}[1]{\text{I}\left[#1\right]}

\newcommand{\idop}{\mathbb{I}}%identity operator

\newcommand{\eball}[2]{B_{#1}\left(#2\right)}%epsilon ball
\newcommand{\polylog}[1]{polylog\left(#1\right)}

\newcommand{\SEP}{\mathbf{SEP}}

\begin{document}

\preprint{APS/123-QED}

\title{Probabilistic state synthesis based on optimal convex approximation}% Force line breaks with \\

\author{Seiseki Akibue}
\email{seiseki.akibue@ntt.com}
\affiliation{%
NTT Communication Science Laboratories, NTT Corporation.\\
3--1, Morinosato-Wakamiya, Atsugi, Kanagawa 243-0198, Japan
}%
\author{Go Kato}%
\affiliation{%
 Advanced ICT Research Institute, NICT.\\
  4--2--1, Nukui-Kitamachi, Koganei, Tokyo 184-8795, Japan
}
\author{Seiichiro Tani}%
\affiliation{%
NTT Communication Science Laboratories, NTT Corporation.\\
3--1, Morinosato-Wakamiya, Atsugi, Kanagawa 243-0198, Japan
}%

\begin{abstract}
 When preparing a pure state with a quantum circuit, there is an unavoidable approximation error due to the compilation error in fault-tolerant implementation. A recently proposed approach called probabilistic state synthesis, where the circuit is probabilistically sampled, is able to reduce the approximation error compared to conventional deterministic synthesis.
In this paper, we demonstrate that the optimal probabilistic synthesis quadratically reduces the approximation error. Moreover, we show that a deterministic synthesis algorithm can be efficiently converted into a probabilistic one that achieves this quadratic error reduction. We also numerically demonstrate how this conversion reduces the $T$-count and analytically prove that this conversion halves an information-theoretic lower bound on the circuit size.
In order to derive these results, we prove general theorems about the optimal convex approximation of a quantum state. Furthermore, we demonstrate that this theorem can be used to analyze an entanglement measure.
\end{abstract}

%Use showkeys class option if keyword
\maketitle

%\tableofcontents
\onecolumngrid
\section{Introduction}
The latest quantum computer applications require various nontrivial quantum states for computation, secure communication, and the fundamental investigation of quantum mechanics. Examples include the ground state (or its approximation) of a Hamiltonian, which is used to compute the ground energy in quantum chemistry \cite{LC19}, a graph state (or its variants \cite{RHBM13, HDERVB06}), which has a wide range of applications such as measurement-based quantum computation \cite{BBRN09}, blind computation \cite{AJE09}, and secret-sharing \cite{DB08}, and data-hiding states, which are utilized for quantum data hiding \cite{DLT02} and the study of local indistinguishability \cite{BDCTEPSW99, MSA09}.
 In addition, quantum linear system solvers \cite{HHL09, CKS17}, which have various applications in machine learning, require a quantum state encoding classical data.

These applications have motivated researchers to optimize a subroutine that synthesizes a target quantum state.
In order to capture the complexity of the state synthesis, there are extensive studies about the size and depth of a circuit consisting of a sequence of $k(\leq2)$-qubit unitary gates needed to generate a target state by applying the circuit to a fixed state $\ket{0}^{\otimes N}$ \cite{MC11, RRIJM16, GH21, ZYY21, XTX22, AYYS22, XGSPS23}.
While these studies focus on the exact synthesis of a target state, a certain level of error is allowed in many quantum information processing protocols and algorithms. 
In practice, we have no choice but to approximately synthesize a target state due to imperfections and discretization when implementing unitary gates in a synthesis circuit.
The imperfection of gates can be almost removed for specific unitary gates, called elementary gates, according to the nature of the system \cite{K03} or the quantum error correction \cite{B15}.
The set of elementary gates is usually a finite set of unitary gates, e.g., Clifford gates (on a constant number of qubits) + $T$ gates, which causes an approximation error when we synthesize a target state since there are infinite quantum states.
We focus on the synthesis of a target state by using a finite number of perfectly implementable elementary gates. 
In this case, the objective of the optimization is reducing the size or depth of a circuit consisting of elementary gates in order to synthesize a target state with a certain level of approximation error. In other words, the objective is to reduce the approximation error within a fixed circuit size or depth.

%Moreover, in practice, we can implement only approximation synthesis due to the error. Even in FTQC era, the compilation error is inevitable.
%In many cases, we assume finite elementary gates are exactly implementable. We would like to minimize the compilation error. or minimize the size. This model has practical significance in the era of fault-tolerant quantum computing (FTQC), as we can arbitrarily reduce noise on each gate according to the quantum error correction \cite{B15} or the nature of the system \cite{K03}. However, the number of almost perfectly implementable gates (called {\it elementary gates}) is usually finite, which results in an approximation error since there are infinite quantum states.
%Thanks to the celebrated Solovay-Kitaev algorithm \cite{KBook}, it is possible to implement an arbitrary unitary transformation as a sequence of {\it universal} elementary gates, e.g., Clifford + $T$ gates, with an arbitrarily small error by increasing the length of the gate sequence. Thus, in the era of FTQC, it is necessary to reduce the size or depth of a circuit consisting of the elementary gates in order to synthesize a target state with a certain level of approximation error. In other words, the objective is to reduce the approximation error within a fixed circuit size or depth.

Unfortunately, a simple volume consideration implies that the size of a circuit required for the approximate synthesis of a quantum state in an $N$-qubit system grows exponentially with $N$. However, it is important to optimize the state synthesis even on a small number of qubits since such small systems are often used repeatedly in quantum cryptography \cite{DB08, DLT02} and metrology \cite{YZ23, JK22} protocols. Such optimization is also beneficial to generate an intermediate quantum state required for synthesizing a state on a large system. 
Recently, theoretical physicists have taken an interest in the minimum circuit size or depth for the state synthesis on large systems due to its nontrivial physical interpretations \cite{FSJRPZJA20, FWNRJ21, I21}, even if it may not be practically implementable.

The final goal of conventional synthesis algorithms is to deterministically find one of the best circuits for the approximation (even if an algorithm \cite{ZYY21} succeeds probabilistically). Thus, the minimum approximation error obtained by such deterministic state synthesis is given by $\min_{x\in X}\trdist{\phi-\hat{\phi}_x}$, where $\phi$ is a target state, $\trdist{\rho-\sigma}$ is the trace distance between two states $\rho$ and $\sigma$, and $X$ is the label set of pure states $\hat{\phi}_x$ generated by circuits $\mathcal{C}_x$ within a given cost, e.g., the circuit size, depth, or number of $T$-gates.

While it makes sense to approximate a target pure state by utilizing an approximated state generated by a single circuit, a recently proposed approach called probabilistic state synthesis probabilistically samples a circuit for the approximation. 
Suppose that the probabilistic algorithm independently samples a circuit $\mathcal{C}_x$ (generating $\hat{\phi}_x$) in accordance with a probability distribution $p(x)$ each time the subroutine synthesizing $\phi$ is called. Then, each generated state is described by a mixed state $\sum_{x}p(x)\hat{\phi}_{x}$.
This can be interpreted as the transition from unitary errors
\if0\footnote{{\red The coherent error usually refers to an error characterized by a unitary transformation taking place on physical qubits. In contrast, we focus on the compilation error causing unitary errors on logical qubits. Since both errors can be described by unitary transformations, we interpret the compilation error as also a coherent error here.}}\fi
 to stochastic errors  \cite{H17, WE16, KN23}, and recent studies have experimentally demonstrated that this transition reduces the approximation error \cite{HNMVMMDSIOHWES21}. 

Despite its importance, the limitation of probabilistic state synthesis, especially the minimum approximation error $\min_p\trdist{\phi-\sum_{x}p(x)\hat{\phi}_x}$, remains unknown, nor is it clear how to find the optimal probability distribution $p$. While a few analytical results are obtained for the case of a qubit state \cite{S17,LHYFH20,ZYY21conv} in the context of the optimal convex approximation of a quantum state, minimax optimization to compute the minimum approximation error makes analyses quite difficult in general.

\subsection{Our contributions}
\begin{figure}[h]
\includegraphics[height=.19\textheight]{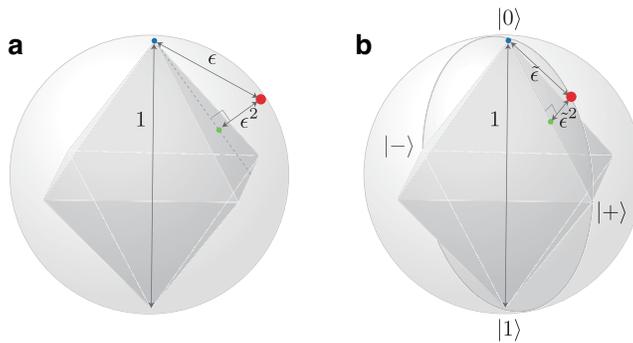}% Here is how to import EPS art
\caption{\label{fig:Bloch} \textbf{Quadratic reduction of the approximation error by using probabilistic synthesis.}
We assume that we can exactly generate an eigenstate $\hat{\phi}_x$ of the Pauli operators, represented by the six extreme points of the octahedron. We represent the Bloch sphere by a sphere with radius $\frac{1}{2}$, where the trace distance between two quantum states equals the Euclidean distance between the corresponding points. (a) We can compute $\min_p\trdist{\phi-\sum_xp(x)\hat{\phi}_x}=\epsilon^2=\frac{1}{2\sqrt{3}}\left(\sqrt{3}-1\right)$ and $\min_x\trdist{\phi-\hat{\phi}_x}=\epsilon$, where $\phi$ is the furthest state from $\{\hat{\phi}_x\}_{x=1}^6$, represented as a large red point. (b) Suppose that the target state is chosen from $S_G:=\{\phi:\ket{\phi}=\cos t\ket{0}+\sin t\ket{1},t\in\rr\}$, represented by a meridian. We can compute $\min_p\trdist{\phi-\sum_xp(x)\hat{\phi}_x}=\tilde{\epsilon}^2=\frac{1}{2}\left(1-\frac{1}{\sqrt{2}}\right)$ and $\min_{x}\trdist{\phi-\hat{\phi}_x}=\tilde{\epsilon}$, where $\phi$ is the furthest state in $S_G$ from $\{\hat{\phi}_x\}_{x=1}^6$, represented as a large red point.}
\end{figure}

Before presenting our results, we provide intuitive examples demonstrating the capability of probabilistic synthesis in Fig.~\ref{fig:Bloch}. As a generalization of the qubit examples, we obtain the fundamental relationship between the minimum approximation errors obtained by the deterministic synthesis and the probabilistic one in the following theorem.

{\bf Theorem \ref{thm:main1}.} (simplified version)
{\it 
 For any subset $\{\hat{\phi}_x\}_{x\in X}$ of pure states, it holds that
\begin{equation}
 \max_{\phi} \min_p\trdist{\phi-\sum_{x\in X}p(x)\hat{\phi}_x}=\max_{\phi}\min_{x\in X}\trdist{\phi-\hat{\phi}_x}^2,
\end{equation}
where the maximization of $\phi$ is taken over the set of pure states.
}

This theorem compares the {\it worst} approximation errors occurring when one synthesizes the target state that is most difficult to approximate by using $\{\hat{\phi}_x\}_{x}$. It implies that the optimal probabilistic synthesis always quadratically reduces the worst approximation error, moreover, it is impossible to further reduce the approximation error.

In many cases, there is no need to synthesize all possible pure states. Instead, it is more useful to understand the limitations of probabilistic synthesis when a target state is chosen from a subset $S_G$ of pure states. As shown in Fig.~\ref{fig:Bloch}(b), we can also anticipate the quadratic error reduction in this scenario. This expectation is confirmed in the comprehensive version of Theorem \ref{thm:main1}, which includes the case of Fig.~\ref{fig:Bloch}(b).

\if0
This theorem compares the {\it worst} approximation errors occurring when one synthesizes the target state in a subset $S_G$ that is most difficult to approximate by using $\{\hat{\phi}_x\}_{x}$, which does not need to be a subset of $S_G$. It implies that the optimal probabilistic synthesis at least quadratically reduces the approximation error for any target state $\phi\in S_G$ compared to the worst approximation error caused by the optimal deterministic synthesis. This theorem holds for various $S_G$ by tailoring $G$. For example, $S_G$ coincides with the set of pure states when $G=\{\idop\}$. In such a case, Theorem \ref{thm:main1} is applicable to any  $\{\hat{\phi}_x\}_x$. When $G=\{\idop,\theta\}$, where $\theta$ represents the complex conjugation with respect to the computational basis, $S_G$ coincides with $\{\cos t\ket{0}+\sin t\ket{1}:t\in\rr\}$, which is generated by an axial rotation $\exp(-it\sigma_Y)$ along a fixed axis from $\ket{0}$.
Such one-parameter pure states, more generally known as conjugation-invariant pure states, are often utilized in the optimal parameter estimation \cite{JK22}. For the last example, $S_G$ coincides with the set of pure states in a subspace $\vv$ or its orthogonal complement $\vv_\bot$ when $G=\{\idop,2\Pi_\vv-\idop\}$, where $\Pi_\vv$ is the Hermitian projector whose range is $\vv$. Preparing a state in a particular subspace is an extensively used subroutine in various quantum information processing tasks.
\fi

The technique used to prove Theorem \ref{thm:main1} is also applicable to analyzing the minimum trace distance between a general mixed state $\rho$ and a convex hull of $\{\hat{\phi}_x\}_{x}$. For example, we can analyze the entanglement measure by setting $\{\hat{\phi}_x\}_{x\in X}$ to be the set of pure product states.
 As a byproduct, we obtain
 \begin{eqnarray}
\label{eq:entconj}
\min_{\sigma\in \SEP}\trdist{\rho_q^{\rm WER}-\sigma}=q-\frac{1}{2},\ \ 
 \min_{\sigma\in \SEP}\trdist{\rho_q^{\rm ISO}-\sigma}=\frac{d^2-1}{d^2}\left(q-\frac{1}{d+1}\right),
\end{eqnarray}
where $\SEP$ represents the set of separable states, $\rho_q^{\rm WER}$ and $\rho_q^{\rm ISO}$ represent the Werner and isotropic state with a parameter $q$, respectively.
These coincide with a conjecture numerically found in \cite{ANT22}. Moreover, we provide alternate succinct proof about a recently identified coincidence between the entanglement measure and coherence measure \cite{JSNCS16}.

We also show an efficient way to convert a deterministic state synthesis algorithm into a probabilistic one that achieves quadratic error reduction. We assume there exists a deterministic state synthesis algorithm $\mathcal{D}$ with

INPUT: a target pure state $\phi$  in a constant number of qubits and target approximation error $\epsilon$,

OUTPUT: circuit $\mathcal{C}_x$ (generating $\hat{\phi}_x$)

\noindent
such that $\trdist{\phi-\hat{\phi}_x}\leq\epsilon$ and a matrix representation of $\hat{\phi}_x$ can be obtained within runtime $\polylog{\frac{1}{\epsilon}}$. 
We can construct $\mathcal{D}$ by combining algorithms to generate an exact synthesis circuit where arbitrary unitary transformations on a constant number of qubits are allowed \cite{MC11, RRIJM16, GH21, ZYY21, XTX22, AYYS22, XGSPS23} with the Solovay-Kitaev algorithm \cite{KBook} to decompose the unitary transformations into a sequence of elementary gates. 
Recent numerical analysis suggests that we could construct better $\mathcal{D}$ that reduces the size of a synthesis circuit by skipping the exact synthesis as an intermediate step \cite{AYYS22, TBNAIM23}.
The efficient conversion is shown in the following theorem.
 
{\bf Theorem \ref{thm:main2}.} (informal version)
{\it 
There exists a probabilistic state synthesis algorithm $\mathcal{P}$ that calls a deterministic state synthesis algorithm $\mathcal{D}$ as an oracle, and has

{\rm INPUT}: a target pure state $\phi$  in a constant number of qubits and target approximation error $\epsilon$

{\rm OUTPUT}: circuit $\mathcal{C}_x$ (generating $\hat{\phi}_x$) sampled in accordance with probability distribution $\hat{p}:\hat{X}\rightarrow[0,1]$

\noindent
such that $\mathcal{P}$ satisfies the following properties:
\begin{itemize}
 \item {\rm Efficiency}: $\mathcal{P}$ calls $\mathcal{D}$ constant times, and runtime of $\mathcal{P}$ is $\polylog{\frac{1}{\epsilon}}$,
 
 \item {\rm Quadratic improvement}: The approximation error $\trdist{\phi-\sum_{x\in \hat{X}}\hat{p}(x)\hat{\phi}_x}$ obtained with this algorithm is upper bounded by $\epsilon^2$, whereas $\min_{x\in \hat{X}}\trdist{\phi-\hat{\phi}_{x}}\leq\epsilon$.
\end{itemize}
}

Since probabilistic state synthesis reduces the approximation error, it also reduces the size of a circuit to approximately generate a target state for a given approximation error. However, the reduction rate depends on the circuit's construction, e.g., what kind of elementary gates and synthesis algorithms are used. 
Since there is an established way to synthesize a single qubit state by using Clifford + $T$ gates, we perform a numerical simulation to demonstrate how the probabilistic synthesis reduces the number of  $T$-gates, called a $T$-count, for a randomly selected target state in $S_G$ defined in Fig.~\ref{fig:Bloch}(b).

\begin{figure}[h]
\includegraphics[height=.045\textheight]{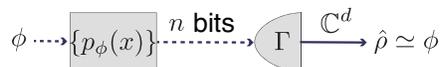}% Here is how to import EPS art
\caption{\label{fig:Cencoding} \textbf{Probabilistic encoding of pure state $\phi$ on a $d$-dimensional system using $n$-bit strings and a decoder $\Gamma$ that generates an approximated state $\hat{\rho}$.} State $\phi$ is probabilistically encoded in label $x$ in a finite set $X$ in accordance with probability distribution $p_\phi:X\rightarrow[0,1]$. As a special case of probabilistic encoding, we also consider deterministic encoding that utilizes probability distribution $p_\phi:X\rightarrow\{0,1\}$. Note that the length of classical bit strings to represent $x\in X$ is given by $n=\lceil\log_2|X|\rceil$.}
\end{figure}

As a rigorous estimation, we also analyze a universal lower bound on the size of synthesis circuits obtained by regarding the circuit as a classical encoding of a pure state, where a description of a circuit $\mathcal{C}_x$ and the state $\hat{\phi}_x$ generated by $\mathcal{C}_x$ correspond to a label encoding a pure state and the reconstructed state by a decoder, respectively. To analyze how probabilistic synthesis reduces this lower bound, we investigate the minimum length of classical bit strings that encodes a pure state $\phi$ so as to approximately reconstruct the original state as shown in Fig.~\ref{fig:Cencoding}. 

We compare two types of encoding: (1) deterministic encoding that associates each $\phi$ to a single label $x$, and (2) probabilistic encoding that associates each $\phi$ to a label $x$ in accordance with a probability distribution $p_\phi(x)$. The decoder $\Gamma$ generates, in general, a mixed state $\hat{\rho}_x$ based on the input label $x$. Thus, the reconstructed state in the deterministic and probabilistic encoding is given by $\hat{\rho}=\hat{\rho}_x$ and $\hat{\rho}=\sum_xp_\phi(x)\hat{\rho}_x$, respectively.
In the following theorem, we show that probabilistic encoding exactly halves the bit length required for deterministic encoding in the asymptotic limits.

{\bf Theorem \ref{thm:main3}.} (simplified version)
{\it 
 Let $n_{\rm det}$ (or $n_{\rm prob}$) be the minimum bit length required for deterministic (or probabilistic) encoding that reconstructs a state $\hat{\rho}$ satisfying $\trdist{\phi-\hat{\rho}}\leq\epsilon$ for any pure state $\phi$ in a $d$-dimensional Hilbert space. Then, it holds that
 \begin{equation}
 \lim_{\epsilon\rightarrow0}\frac{n_{\rm prob}}{n_{\rm det}}= \lim_{d\rightarrow\infty}\frac{n_{\rm prob}}{n_{\rm det}}=\frac{1}{2}.
\end{equation}
}

\if0
 While this theorem implies the half reduction of a lower bound on the size of a synthesis circuit, the same reduction of the minimum number of $T$ gates is possible if we assume a conjecture about the minimum number \cite{RS16}.
\fi

\subsection{Technical Outline}

Although several probabilistic synthesis methods suggest that the approximation error can be reduced from $\epsilon$ into $O(\epsilon^2)$ \cite{C17, H17, WE16, HNMVMMDSIOHWES21, KLMPP22, KN23}, these methods are not applicable for analyzing the achievable minimum approximation error. This is mainly because the prior research relies on the first-order approximation to show the error reduction, which provides little information about the lower bound on the error reduction.
The achievable minimum approximation error for the probabilistic unitary synthesis has been obtained by us \cite{SGS23}.
However, this result cannot be directly applied to the state synthesis since the generated state in state synthesis is obtained by applying a gate sequence to a fixed input state while the approximation error in unitary synthesis is quantified for the worst input state. Moreover, a target state could be approximated by probabilistically mixing two unitary transformations whose behaviors are totally different, except for the fixed input state.

In the proof of Theorem \ref{thm:main1}, we analyze the minimum approximation error
\begin{equation}
 \min_p\trdist{\phi-\sum_{x}p(x)\hat{\phi}_x}=\min_p\max_{0\leq M\leq\idop}\tr{M(\phi-\sum_{x}p(x)\hat{\phi}_x)},
\end{equation}
which contains minimax optimization by definition. The main tool for the analysis is the strong duality of semidefinite programming. This enables us to formulate the minimum approximation error as a semidefinite program (SDP). 
Moreover, we show that the SDP can be dramatically simplified when both $\phi$ and $\{\hat{\phi}_x\}_x$ exhibit symmetry.
As discussed in the previous subsection, these techniques can be utilized to analyze the minimum trace distance between a general mixed state and a convex set, such as the set of separable states.

The reformulation of the minimum approximation error as an SDP enables us to compute the optimal probability distribution to achieve it efficiently. By using Theorem \ref{thm:main1}, we can verify that by solving this SDP with $\{\hat{\phi}_x\}_{x\in X}$ satisfying $\max_\phi\min_{x\in X}\trdist{\phi-\hat{\phi}_x}\leq\epsilon$, which is called an $\epsilon$-covering, we obtain a probability distribution $\hat{p}$ that achieves quadratic reduction of the approximation error, i.e., $\trdist{\phi-\sum_{x\in X}\hat{p}(x)\hat{\phi}_x}\leq\epsilon^2$. However, the size of this SDP is too large to achieve the efficiency shown in Theorem \ref{thm:main2}, since the size $|X|$ of the $\epsilon$-covering is $\left(\frac{1}{\epsilon}\right)^{\Omega(1)}$. This problem can be resolved by proving that any $\hat{\phi}_x$ in the support of the optimal probability distribution in the minimum approximation error is close to $\phi$; more precisely, $\trdist{\phi-\hat{\phi}_x}\leq2\epsilon$. This enables us to construct a modified SDP whose size is independent of $\epsilon$.

Theorem \ref{thm:main3} is obtained by combining Theorem \ref{thm:main1} with the estimation of the minimum size of the $\epsilon$-covering. Due to its prominent role in algorithm design and asymptotic geometric analysis, the order of the minimum size of the $\epsilon$-covering has been well-studied \cite{HLW06, GT13, ABmeetBanach}. However, to obtain Theorem \ref{thm:main3}, we precisely analyze the constant factor in the order, which refines the previous estimations \cite{HLW06, GT13}.

\if0
\subsection{Organization}
Section \ref{sec:preliminaries} of this paper introduces the basic notations in quantum information theory and the antiunitary operator, which we use to represent symmetry. In Section \ref{sec:quad}, we show the quadratic reduction of approximation error by probabilistic synthesis. In
Section \ref{sec:PSSalgorithm}, we show an efficient conversion from a deterministic synthesis algorithm into a probabilistic one.
In Section \ref{sec:Nsimulation}, we perform a numerical simulation to demonstrate how this conversion reduces a $T$-count.
In Section \ref{sec:Cbit}, we show that probabilistic encoding asymptotically halves the length of a classical bit string required to approximately encode a pure state.
As a byproduct of our technique, we provide exact formulas and alternate concise proof about an entanglement measure in Section \ref{sec:application}.
We conclude Section \ref{sec:conclusion} with a brief summary.
\fi

\section{Results}
\subsection{Preliminaries}
\label{sec:preliminaries}
We consider only finite-dimensional Hilbert spaces in this paper. The two-dimensional Hilbert space $\cdim{2}$ is called a qubit.
$\linop{\hh}$ and $\pos{\hh}$ represent the set of linear operators and positive semidefinite operators on Hilbert space $\hh$, respectively. 
$\idop\in\pos{\hh}$ represents the identity operator.
For Hermitian operators $A$ and $B$ on $\hh$, $A\geq B$ represents $A-B\in\pos{\hh}$, and $A>B$ means $A-B$ is positive definite.
$\dop{\hh}:=\left\{\rho\in\pos{\hh}:\tr{\rho}=1\right\}$ and $\puredop{\hh}:=\left\{\rho\in\dop{\hh}:\tr{\rho^2}=1\right\}$ represent the set of quantum states and pure states, respectively. Pure state $\phi\in\puredop{\hh}$ is sometimes alternatively represented by complex unit vector $\ket{\phi}\in\hh$ satisfying $\phi=\ketbra{\phi}$. 
\if0
We denote Pauli operators by $\sigma_X$, $\sigma_Y$ and $\sigma_Z$ whose matrix representations with respect to the computational basis are given by
\begin{equation}
 \sigma_X=
 \begin{pmatrix}
 0&1\\
 1&0
\end{pmatrix},
 \sigma_Y=
 \begin{pmatrix}
 0&-i\\
 i&0
\end{pmatrix},
 \sigma_Z=
 \begin{pmatrix}
 1&0\\
 0&-1
\end{pmatrix}.
\end{equation}
\fi

The trace distance $\trdist{\rho-\sigma}$ of two quantum states $\rho,\sigma\in\dop{\hh}$ is defined as $\trdist{M}:=\frac{1}{2}\tr{\sqrt{MM^\dag}}$ for $M\in\linop{\hh}$. It represents the maximum total variation distance between probability distributions obtained by measurements performed on two quantum states. Thus, it satisfies $\trdist{\rho-\sigma}=\max_{0\leq M\leq\idop}\tr{M(\rho-\sigma)}$.
A similar notion measuring the distinguishability of $\rho$ and $\sigma$ is the fidelity function, defined by $\fidelity{\rho}{\sigma}:=\max\tr{\Phi^\rho\Phi^\sigma}$, where $\Phi^\rho\in\puredop{\hh\otimes\hh'}$ is a purification of $\rho$, i.e., $\rho=\ptr{\hh'}{\Phi^\rho}$, and the maximization is taken over all the purifications. Fuchs-van de Graaf inequalities \cite{FG99} provide relationships between the two measures with respect to the distinguishability as follows:
\begin{equation}
\label{ineq:FG}
 1-\sqrt{\fidelity{\rho}{\sigma}}\leq\trdist{\rho-\sigma}\leq\sqrt{1-\fidelity{\rho}{\sigma}}
\end{equation}
holds for any states $\rho,\sigma\in\dop{\hh}$, where the equality of the right inequality holds when $\rho$ and $\sigma$ are pure. 

An operator $A:\hh\rightarrow\hh$ is called antilinear if it satisfies $A(\alpha\ket{\phi}+\beta\ket{\psi})=\alpha^*A\ket{\phi}+\beta^*A\ket{\psi}$, where $\alpha^*$ represents the complex conjugate of $\alpha\in\cc$. The Hermitian adjoint $A^\dag$ of an antilinear operator $A$ is defined by $\bra{\psi}A^\dag\ket{\phi}=\bra{\phi}A\ket{\psi}$. An antilinear operator $U$ is called antiunitary if it satisfies $U^\dag U=\idop$.
An antiunitary operator $\Theta$ is called a conjugation if it satisfies $\Theta^\dag=\Theta$.
An example of a conjugation is the complex conjugation $\theta$ with respect to the computational basis.
Note that for Hermitian operators $M_1$ and $M_2$ and an antilinear operator $A$, the cyclic property $\tr{M_1AM_2A^\dag}=\tr{A^\dag M_1 AM_2}$ of the trace holds.

\subsection{Quadratic reduction of approximation error}
\label{sec:quad}
We first show the lower bound of the approximation error obtained by the optimal probabilistic mixture in the following lemma.
\begin{lemma}
\label{lemma:lowerbound}
For a finite set $\{\hat{\phi}_x\}_{x\in X}\subseteq\puredop{\hh}$ of pure states and a pure state $\phi\in\puredop{\hh}$, it holds that
 \begin{equation}
 \label{ineq:lowerbound}
 \min_p\trdist{\phi-\sum_{x\in X}p(x)\hat{\phi}_x}\geq\min_{x\in X}\trdist{\phi-\hat{\phi}_x}^2.
\end{equation}
\end{lemma}
\begin{proof}
Let $p$ minimize the left-hand side of Eq.~\eqref{ineq:lowerbound}. The following calculation completes the proof.
 \begin{eqnarray}
 (L.H.S.)\geq \left(1-\sum_{x\in X}p(x)\tr{\phi\hat{\phi}_x}\right)\geq\min_{x\in X}\left(1-\fidelity{\phi}{\hat{\phi}_x}\right)=(R.H.S.),
\end{eqnarray}
where we use $\trdist{\rho-\sigma}\geq\max_{\phi\in\puredop{\hh}}\tr{\phi(\rho-\sigma)}$ in the first inequality and use the right equality in Ineq.~\eqref{ineq:FG} in the last equality.
\end{proof}

This lemma shows that the reduction rate of the approximation error by using probabilistic synthesis is, at best, quadratic. However, the two examples given in Fig.~\ref{fig:Bloch} indicate that a precisely quadratic reduction is possible if we consider the worst approximation error occurring when we synthesize the target state that is most difficult to approximate in a particular subset $S_G$ of states.
 To achieve the quadratic reduction, it is important to carefully select $S_G$. We use group symmetries in the following lemma to characterize $S_G$ and prove the quadratic reduction. This characterization also makes it easier to apply this lemma to various settings in the state synthesis.

\begin{lemma}
\label{lemma:worstcase}
  Let $X$ be a finite set, $G$ be a finite subgroup of unitary and antiunitary operators, and $S_G:=\{\phi\in\puredop{\hh}:\forall U\in G,[U,\phi]=0\}$ be the set of pure states invariant under the action of $G$. 
  If a set $\{\hat{\phi}_x\in\puredop{\hh}\}_{x\in X}$ of pure states is invariant under the action of $G$, i.e., $\{\hat{\phi}_x\}_{x\in X}=\{U\hat{\phi}_x U^\dag\}_{x\in X}$ for all $U\in G$, it holds that
\begin{equation}
 \max_{\phi\in S_G} \min_p\trdist{\phi-\sum_{x\in X}p(x)\hat{\phi}_x}=\max_{\phi\in S_G}\min_{x\in X}\trdist{\phi-\hat{\phi}_x}^2.
\end{equation}
\end{lemma}

Lemma \ref{lemma:worstcase} is a direct consequence of the following lemma for computing the minimum trace distance between a mixed state and a convex subset of mixed states.

\begin{lemma}
\label{lemma:mindist}
 Let $X$ be a finite set and $G$ be a finite subgroup of unitary and antiunitary operators.
Let $P_G$ be the set of positive semidefinite operators invariant under the action of $G$, i.e., $P_G:=\{P\in\pos{\hh}:\forall U\in G,[U,P]=0\}$.
  If $\rho\in P_G\cap\dop{\hh}$ and a set $\{\hat{\rho}_x\in\dop{\hh}\}_{x\in X}$ of mixed states is invariant under the action of $G$, i.e., $\{\hat{\rho}_x\}_{x\in X}=\{U\hat{\rho}_x U^\dag\}_{x\in X}$ for all $U\in G$, it holds that
\begin{equation}
\label{eq:witness1}
 \min_p\trdist{\rho-\sum_{x\in X}p(x)\hat{\rho}_x}=\max_{\substack{0\leq M\leq\idop\\ M\in P_G}}\left(\tr{M\rho}-\max_{x\in X}\tr{M\hat{\rho}_x}\right),
\end{equation}
where the minimization is taken over a probability distribution $p$ over $X$.
In particular, when $\rho$ is a pure state $\phi$, it holds that
 \begin{equation}
\label{eq:witness2}
 \min_p\trdist{\phi-\sum_{x\in X}p(x)\hat{\rho}_x}=\max_{\psi\in P_G\cap\puredop{\hh}}\left(\tr{\psi\phi}-\max_{x\in X}\tr{\psi\hat{\rho}_x}\right).
\end{equation}
\end{lemma}

\begin{proof}
We start from a mixed state $\rho$. By using the minimax theorem, we obtain
\begin{eqnarray}
 (L.H.S.\  of\  Eq.~\eqref{eq:witness1})&=&\min_p\max_{0\leq M\leq\idop}\left(\tr{M\rho}-\sum_{x\in X}p(x)\tr{M\hat{\rho}_x}\right)\\
 &=&\max_{0\leq M\leq\idop}\min_p\left(\tr{M\rho}-\sum_{x\in X}p(x)\tr{M\hat{\rho}_x}\right)\\
\label{eq:subspacelemma1}
 &=&\max_{0\leq M\leq\idop}\left(\tr{M\rho}-\max_{x\in X}\tr{M\hat{\rho}_x}\right).
\end{eqnarray}
This proves $(L.H.S.)\geq(R.H.S.)$.
Let $M$ maximize Eq.~\eqref{eq:subspacelemma1}. Due to the invariance of $\rho$ and $\{\hat{\rho}_x\}_x$ under the action of $G$, we can verify that $U^\dag MU$ also maximizes Eq.~\eqref{eq:subspacelemma1}. By defining $\hat{M}=\frac{1}{|G|}\sum_{U\in G}U^\dag MU$, we obtain
\begin{eqnarray}
 (R.H.S.\  of\  Eq.~\eqref{eq:witness1})&\geq&\tr{\hat{M}\rho}-\max_{x\in X}\tr{\hat{M}\hat{\rho}_x}
 =\tr{M\rho}-\max_{x\in X}\left(\frac{1}{|G|}\sum_{U\in G}\tr{MU\hat{\rho}_xU^\dag}\right)\nonumber\\
&\geq&\tr{M\rho}-\frac{1}{|G|}\sum_{U\in G}\max_{x\in X}\tr{MU\hat{\rho}_xU^\dag}\nonumber\\
&=&\tr{M\rho}-\max_{x\in X}\tr{M\hat{\rho}_x}=(L.H.S.\  of\  Eq.~\eqref{eq:witness1}),
\end{eqnarray}
where we use Eq.~\eqref{eq:subspacelemma1} in the last equality.

When $\rho$ is a pure state $\phi$, we can derive 
\begin{equation}
\label{eq:subspacelemma2}
  (L.H.S.\  of\  Eq.~\eqref{eq:witness2})=\max_{\sigma\in P_G\cap\dop{\hh}}\left(\tr{\sigma\phi}-\max_{x\in X}\tr{\sigma\hat{\rho}_x}\right)
\end{equation}
by the same argument starting from
\begin{equation}
 (L.H.S.\  of\  Eq.~\eqref{eq:witness2})=\min_p\max_{\sigma\in\dop{\hh}}\tr{\sigma\left(\phi-\sum_{x\in X}p(x)\hat{\rho}_x\right)}=\max_{\sigma\in\dop{\hh}}\left(\tr{\sigma\phi}-\max_{x\in X}\tr{\sigma\hat{\rho}_x}\right),
\end{equation}
where we use the fact that the dimension of the eigenspace of $\phi-\sum_{x\in X}p(x)\hat{\rho}_x$ associated with positive eigenvalues is zero or one in the first equality, and use the minimax theorem in the second equality. We complete the proof of Eq.~\eqref{eq:witness2} by using the following observation: When $(L.H.S.\  of\  Eq.~\eqref{eq:witness2})=0$, Eq.~\eqref{eq:witness2} holds since there exists $x\in X$ such that $\hat{\rho}_x=\phi$. When $(L.H.S.\  of\  Eq.~\eqref{eq:witness2})>0$, $\sigma$ maximizing Eq.~\eqref{eq:subspacelemma2} is a pure state. For if $\sigma$ with $\lpnorm{\infty}{\sigma}<1$ maximizes Eq.~\eqref{eq:subspacelemma2}, we can show a contradiction by setting $\rho=\phi$ and $M=\frac{\sigma}{\lpnorm{\infty}{\sigma}}$ in Eq.~\eqref{eq:subspacelemma1}.

\end{proof}

\begin{proof}[Proof of Lemma \ref{lemma:worstcase}]
By setting $\hat{\rho}_x$ in Eq.~\eqref{eq:witness2} to be $\hat{\phi}_x$, we obtain
\begin{eqnarray}
 \max_{\phi\in S_G} \min_p\trdist{\phi-\sum_{x\in X}p(x)\hat{\phi}_x}&=&\max_{\psi\in S_G}\left(\max_{\phi\in S_G}\tr{\psi\phi}-\max_{x\in X}\tr{\psi\hat{\phi}_x}\right)\nonumber\\
 &=&1-\min_{\psi\in S_G}\max_{x\in X}\tr{\psi\hat{\phi}_x}=\max_{\psi\in S_G}\min_{x\in X}\trdist{\psi-\hat{\phi}_x}^2,
 \end{eqnarray}
 where we use the right equality in Ineq.~\eqref{ineq:FG} in the last equality.
\end{proof}

As consequences of Lemma \ref{lemma:worstcase} or Lemma \ref{lemma:mindist}, we obtain the following implications.
\begin{enumerate}
 \item When $G=\{\idop\}$, we obtain $S_G=\puredop{\hh}$. This case is applicable to any $\{\hat{\phi}_x\}_{x\in X}$ and proves the quadratic reduction of the approximation error given in Fig.~\ref{fig:Bloch}(a).
 
  \item When $G=\{\idop,\theta\}$ with the complex conjugation $\theta$, we obtain $S_G=\{\phi\in\puredop{\cdim{2}}:\ket{\phi}=\cos t\ket{0}+\sin t\ket{1},t\in\rr\}$. In this case, the quadratic reduction of the worst approximation error occurring when we synthesize a target state in $S_G$ is possible if $\{\hat{\phi}_x\}_x$ is reflection-symmetric with respect to the XZ-plane in the Bloch representation. This proves the quadratic reduction of the approximation error given in Fig.~\ref{fig:Bloch}(b). In general, conjugation-invariant pure states are often utilized in the optimal parameter estimation \cite{JK22}.

  \item When $G=\{\idop,2\Pi-\idop\}$ with Hermitian projector $\Pi$ whose range is $\vv$, $S_G=\{\phi\in\puredop{\hh}:\ket{\phi}\in\vv\vee\ket{\phi}\in\vv_{\bot}\}$. In this case, the quadratic reduction of the worst approximation error occurring when we synthesize a target state in $\vv$ is possible if $\{\hat{\phi}_x\}_x$ is reflection-symmetric under the action of $2\Pi-\idop$. This is because 
 \begin{eqnarray}
\max_{\ket{\phi}\in\vv}\min_p\trdist{\phi-\sum_{x\in X}p(x)\hat{\phi}_x}&=&\max_{\psi\in S_G}\left(\max_{\ket{\phi}\in \vv}\tr{\psi\phi}-\max_{x\in X}\tr{\psi\hat{\phi}_x}\right)\\
&=&\max_{\ket{\psi}\in \vv}\left(\max_{\ket{\phi}\in \vv}\tr{\psi\phi}-\max_{x\in X}\tr{\psi\hat{\phi}_x}\right)\\
&=&1-\min_{\ket{\psi}\in\vv}\max_{x\in X}\tr{\psi\hat{\phi}_x}=\max_{\ket{\phi}\in\vv}\min_{x\in X}\trdist{\phi-\hat{\phi}_x}^2,
\end{eqnarray}
where we use Eq.~\eqref{eq:witness2} in the first equation.
In general, preparing a state in a particular subspace is a widely used subroutine in various quantum information processing tasks.
\end{enumerate}

We obtain the following theorem as a summary of Lemmas \ref{lemma:lowerbound} and \ref{lemma:worstcase}.
\begin{theorem}
\label{thm:main1}
 Let $X$ be a finite set, $G$ be a finite subgroup of unitary and antiunitary operators, and $S_G:=\{\phi\in\puredop{\hh}:\forall U\in G,[U,\phi]=0\}$ be the set of pure states invariant under the action of $G$.
  If $\phi\in S_G$ and $\{\hat{\phi}_x\in\puredop{\hh}\}_{x\in X}=\{U\hat{\phi}_x U^\dag\}_{x\in X}$ for all $U\in G$, it holds that
 \begin{equation}
 \epsilon_\phi^2\leq \min_p\trdist{\phi-\sum_{x\in X}p(x)\hat{\phi}_x}\leq\epsilon_G^2\ \ \ {\rm with}\ \epsilon_\phi=\min_{x\in X}\trdist{\phi-\hat{\phi}_x},\ \epsilon_G=\max_{\psi\in S_G}\min_{x\in X}\trdist{\psi-\hat{\phi}_x}.
\end{equation}
 \end{theorem}

This theorem indicates that by using mixed states, we can reduce the approximation error with respect to the trace distance. 
When attempting to estimate the expectation value $\tr{O\phi}$ of an observable $O$ for $\phi$, this theorem implies that the bias of the expectation value can be reduced by using $\sum_{x\in X}p(x)\hat{\phi}_x$ instead of using $\hat{\phi}_x$ as a substitute of $\phi$.

\subsection{Efficient probabilistic state synthesis algorithm}
\label{sec:PSSalgorithm}
 In this section, we present an efficient method for converting any deterministic state synthesis algorithm, denoted as $\mathcal{D}$, into a probabilistic one. If it takes $\polylog{\frac{1}{\epsilon}}$-time for $\mathcal{D}$ \footnote{A deterministic synthesis based on the Solovay-Kitaev algorithm is an example of such $\mathcal{D}$.} to achieve an approximation error $\epsilon$ with an $l(\epsilon)$-size circuit, then our method allows us to construct a probabilistic synthesis algorithm that achieves an approximation error $\epsilon^2$ by sampling $l(\epsilon)$-size circuits, with a total runtime of $\polylog{\frac{1}{\epsilon}}$. 

Note that our method assumes the target state is taken from a constant-dimensional Hilbert space. As mentioned in the introduction, constant-qubits states are commonly utilized in quantum cryptography and metrology protocols. Although the existence of highly complex pure states results in an exponential runtime with respect to the number of qubits for any state synthesis algorithms, we discuss the potential of probabilistic state synthesis for a high dimensional system in Appendix \ref{sec:high dimension}.
Our conversion is based on the following proposition and lemma.

\begin{proposition}
\label{prop:SDP}
Let $\rho$ and $\{\hat{\rho}_x\}_{x\in X}$ be a target mixed state and a finite set of mixed states in $\dop{\hh}$, respectively. Then, distance $\min_{p}\trdist{\rho-\sum_{x\in X}p(x)\hat{\rho}_x}$ and the optimal probability distribution $\{p(x)\}_{x\in X}$, which minimizes the distance, can be computed with the following SDP:
\begin{equation}
\label{eq:SDP}
\begin{tabular}{rlcrl}
\multicolumn{2}{c}{\underline{{\rm Primal problem}}} &\ \ \ \ \ \ \ \ \ \ \ \ \ \  
 &\multicolumn{2}{c}{\underline{{\rm Dual problem}}}\\
{\rm maximize:}&$\tr{M\rho}-t$&&{\rm minimize:}&$\tr{Y}$ \\
{\rm subject to:}& $0\leq M\leq\idop$,&&
{\rm subject to:}& $Y\geq0\wedge Y\geq \rho-\sum_{x\in X}p(x)\hat{\rho}_x$,\\
&$\forall x\in X,\tr{M\hat{\rho}_x}\leq t$.&&&$\forall x\in X,p(x)\geq0$,\\
&&&&$\sum_{x\in X}p(x)\leq 1$.
\end{tabular} 
\end{equation}
Note that the strong duality holds in this SDP, i.e., the optimum primal and dual values are equal.

\end{proposition}
\begin{proof}
Recall that for two states $\rho$ and $\sigma$, $\trdist{\rho-\sigma}$ can be computed by the following SDP:
\begin{center}
 \begin{tabular}{rlcrl}
\multicolumn{2}{c}{\underline{Primal problem}} &\ \ \ \ \ \ \ \ \ \ \ \ \ \ \ \ \ \ \ \ \ \ \ \ \ \ \ \ \ \ \ \  & \multicolumn{2}{c}{\underline{Dual problem}}\\
maximize: & $\tr{M(\rho-\sigma)}$ && minimize:& $\tr{Y}$\\
subject to: & $0\leq M\leq\idop$, && subject to: & $Y\geq0\wedge Y\geq \rho-\sigma$.\\
\end{tabular}
\end{center}

A formal SDP and the verification of the strong duality are provided in Appendix \ref{appendix:SDP}.

By extending the dual problem of this SDP to include the minimization of probability distribution $\{p(x)\}_{x\in X}$, we obtain Eq.~\eqref{eq:SDP}. Note that the last condition $\sum_{x\in X}p(x)\leq 1$ in the dual problem is different from the condition $\sum_{x\in X}p(x)=1$ of a probability distribution; however, the optimum dual value can be achieved under the latter condition.
Again, a formal SDP and the verification of the strong duality are provided in Appendix \ref{appendix:SDP}.
\end{proof}

\begin{lemma}
\label{lemma:support}
Let $G$ be a finite subgroup of unitary and antiunitary operators, and $S_G:=\{\phi\in\puredop{\hh}:\forall U\in G,[U,\phi]=0\}$ be the set of pure states invariant under the action of $G$. 
For a positive number $\epsilon>0$, if $\phi\in S_G$ and $\{\hat{\phi}_x\}_{x\in X}$ is a finite $\epsilon$-covering of $S_G$ that is invariant under the action of $G$, i.e., $\max_{\psi\in S_G}\min_{x\in X}\trdist{\psi-\hat{\phi}_x}\leq\epsilon$ and $\{\hat{\phi}_x\}_{x\in X}=\{U\hat{\phi}_xU^\dag\}_{x\in X}$ for all $U\in G$, then
\begin{equation}
\label{eq:support}
\min_{p}\trdist{\phi-\sum_{x\in X}p(x)\hat{\phi}_x}=\min_{\hat{p}}\trdist{\phi-\sum_{x\in \hat{X}}\hat{p}(x)\hat{\phi}_x}
\end{equation}
holds, where $\hat{X}:=\{x\in X:\trdist{\phi-\hat{\phi}_x}\leq2\epsilon\}$ and the minimization of $p$ and $\hat{p}$ are taken over probability distributions over $X$ and $\hat{X}$, respectively.
\end{lemma}
To understand this lemma, it is helpful to refer to the examples shown in Fig.~\ref{fig:Bloch}. If the goal is to optimally approximate a target state $\phi$ depicted by the red point in (a) (or (b)), it is sufficient to mix three (or two) Pauli eigenstates that are $2\epsilon$ (or $2\tilde{\epsilon}$) close to $\phi$. This fact is shown to be true for any target state in this lemma, and its proof can be found in Appendix \ref{sec:proof_support} as it involves technical details.

By combining Proposition \ref{prop:SDP} and Lemma \ref{lemma:support}, we can efficiently convert a deterministic state synthesis algorithm into a probabilistic one. We assume there exists a deterministic state synthesis algorithm $\mathcal{D}$ with

INPUT: a target pure state $\phi\in S_G$  in a constant-dimensional Hilbert space and a target approximation error $\epsilon\in\left(0,1\right)$

OUTPUT: a set $\{\mathcal{C}_x^{(U)}\}_{U\in G}$ of circuits (generating $U\hat{\phi}_xU^\dag$)

such that $\trdist{\phi-\hat{\phi}_x}\leq\epsilon$ and a matrix representation of $U\hat{\phi}_xU^\dag$ can be obtained within runtime $\polylog{\frac{1}{\epsilon}}$, where $G$ is a finite subgroup of unitary and antiunitary operators and $S_G$ is the set of pure states invariant under the action of $G$.

\if0
When $G=\{\idop\}$, concrete constructions of $\mathcal{D}$ are given in the introduction.
When $G=\{\idop,\theta\}$ with the complex conjugation $\theta$ and every output circuit of $\mathcal{D}$ consists of the Hadamard gate, controlled-NOT, $T$ gate and $S(=T^2)$ gate, we can construct $\mathcal{C}_x^{(\theta)}$ by replacing a gate $T$ and $S$ by a gate sequence $S^3T(=T^*)$ and $S^3(=S^*)$, respectively, in $\mathcal{C}_x^{(\idop)}$ without increasing the number of $T$ gates.
\fi

\begin{theorem}
\label{thm:main2}
For a given gate set, there exists a probabilistic state synthesis algorithm $\mathcal{P}$ that calls a deterministic synthesis algorithm $\mathcal{D}$ as an oracle, and has

{\rm INPUT}: a target pure state $\phi\in S_G$ in a constant-dimensional Hilbert space, a target approximation error $\epsilon\in\left(0,1\right)$, and precision $\delta\in\left(0,1\right)$

{\rm OUTPUT}: circuit $\mathcal{C}_x$ (generating $\hat{\rho}_x$) sampled from a set $\hat{X}$ in accordance with probability distribution $\hat{p}:\hat{X}\rightarrow[0,1]$

\noindent
such that $\mathcal{P}$ satisfies the following properties:
\begin{itemize}
 \item {\rm Efficiency}: $\mathcal{P}$ calls $\mathcal{D}$ a constant number of times, and runtime of $\mathcal{P}$ is $poly\left(\log\left(\frac{1}{\epsilon}\right),\log\left(\frac{1}{\delta}\right)\right)$,
 
 \item {\rm Quadratic improvement}: The approximation error $\trdist{\phi-\sum_{x\in\hat{X}}\hat{p}(x)\hat{\rho}_x}$ obtained by $\mathcal{P}$ is upper bounded by $\epsilon^2+\delta$, whereas $\min_{x\in\hat{X}}\trdist{\phi-\hat{\rho}_x}\leq\epsilon$.
 
\end{itemize}
\end{theorem}
\begin{proof}
 In the following, we explicitly construct the algorithm.

%\noindent \textbf{Efficient probabilistic state synthesis algorithm}
\begin{enumerate}
\item Set free parameters $c>0$ and $c'>0$ satisfying $c+c'\leq1$.

 \item Generate a list $\{\phi_x\}_{x\in \tilde{X}}\subseteq S_G$ such that for any $\psi\in S_G$, $\min_{x\in \tilde{X}}\trdist{\psi-\phi_x}\leq c\epsilon$ if $\trdist{\phi-\psi}\leq 2\epsilon$. That is, $\{\phi_x\}_{x\in \tilde{X}}$ is a $(c\epsilon)$-covering of $\{\psi\in S_G: \trdist{\phi-\psi}\leq 2\epsilon\}$.
  
 \item Call $\mathcal{D}$ to find $\mathcal{C}_x^{(U)}$ generating $U\hat{\phi}_x U^\dag$ such that $\trdist{\phi_x-\hat{\phi}_x}\leq c'\epsilon$ for all $x\in\tilde{X}$ and all $U\in G$.
 
 \item Numerically solve the SDP shown in Proposition \ref{prop:SDP} by setting $\rho=\phi$ and $\{\hat{\rho}_x\}_{x\in\hat{X}}=\{U\hat{\phi}_xU^\dag\}_{x\in\tilde{X},U\in G}$ and obtain a probability distribution $\hat{p}$, which causes the approximation error $\delta$-close to $\min_p\trdist{\phi-\sum_{x\in\hat{X}}p(x)\hat{\rho}_x}$.
 
 \item Sample $\mathcal{C}_x^{(U)}$ in accordance with $\hat{p}$, whose domain is $\hat{X}=\tilde{X}\times G$.
\end{enumerate}
The two properties can be verified as follows:
\begin{itemize}
 \item {\it Efficiency}: We can verify that all steps of the algorithm take $poly\left(\log\left(\frac{1}{\epsilon}\right),\log\left(\frac{1}{\delta}\right)\right)$-time by using the following observations: We can construct a list $\{\phi_x\}_{x\in\tilde{X}}$ whose size is independent to $\epsilon$. From the assumption on $\mathcal{D}$, we can also obtain a list of matrix representations of $\{U\phi_xU^\dag\}_{x\in\tilde{X},U\in G}$ within $\polylog{\frac{1}{\epsilon}}$-time. 
The ellipsoid method guarantees that the optimal value of our SDP can be computed in $poly\left(\log\left(\frac{1}{\epsilon}\right),\log\left(\frac{1}{\delta}\right)\right)$-time within an approximation error $\delta$ \cite{L03}.
  
 \item {\it Quadratic improvement}: The minimum approximation error $\min_p\trdist{\phi-\sum_{x\in\hat{X}}p(x)\hat{\rho}_x}$ is at most $\epsilon^2$ since $\{\hat{\rho}_x\}_{x\in\hat{X}}$ is a subset of an $\epsilon$-covering $\{\hat{\rho}_x\}_{x\in\hat{X}}\cup\{\psi_y\}_y$ of $S_G$, where $\{\psi_y\in S_G\}_y$ is a finite $\epsilon$-covering of $\{\psi\in S_G:\trdist{\phi-\psi}>2\epsilon\}$ and $\trdist{\phi-\psi_y}> 2\epsilon$ for any $y$, $\{\hat{\rho}_x\}_{x\in\hat{X}}\cup\{\psi_y\}_y$ is invariant under the action of $G$, and we can thus apply Theorem \ref{thm:main1} and Lemma \ref{lemma:support}.
 
\end{itemize}

\end{proof}

While this theorem assumes the dimension $d$ of the Hilbert space is constant, we can also provide an estimation of the runtime of $\mathcal{P}$ when $d$ grows. The runtime varies depending on the symmetry $G$ that target states possess (see Appendix \ref{sec:high dimension}). In the worst case where target states have no common symmetry, i.e., $G=\{\idop\}$, the size of $\hat{X}$ will be $|\hat{X}|=poly(\exp(d))$. In this case, we can provide the upper bound on the runtime of $\mathcal{P}$ as $poly\left(\log\left(\frac{1}{\epsilon}\right),\log\left(\frac{1}{\delta}\right),\exp(d)\right)$-time, based on the proof of Theorem \ref{thm:main2}.

\subsection{Numerical simulation of $T$-count reduction}
\label{sec:Nsimulation}
In this section, we demonstrate how Theorem \ref{thm:main2}'s probabilistic synthesis algorithm can reduce the $T$-count through numerical simulation.
We select a target state $\phi$ from $S_G=\{\phi\in\puredop{\cdim{2}}:\ket{\phi}=\cos t\ket{0}+\sin t\ket{1},t\in\rr\}$, as shown in Fig.~\ref{fig:Bloch}(b). 
Recall that $S_G$ consists of $G$-invariant pure states, where $G=\{\idop,\theta\}$ with the complex conjugation $\theta$.

\begin{figure}[h]
\includegraphics[height=.35\textheight]{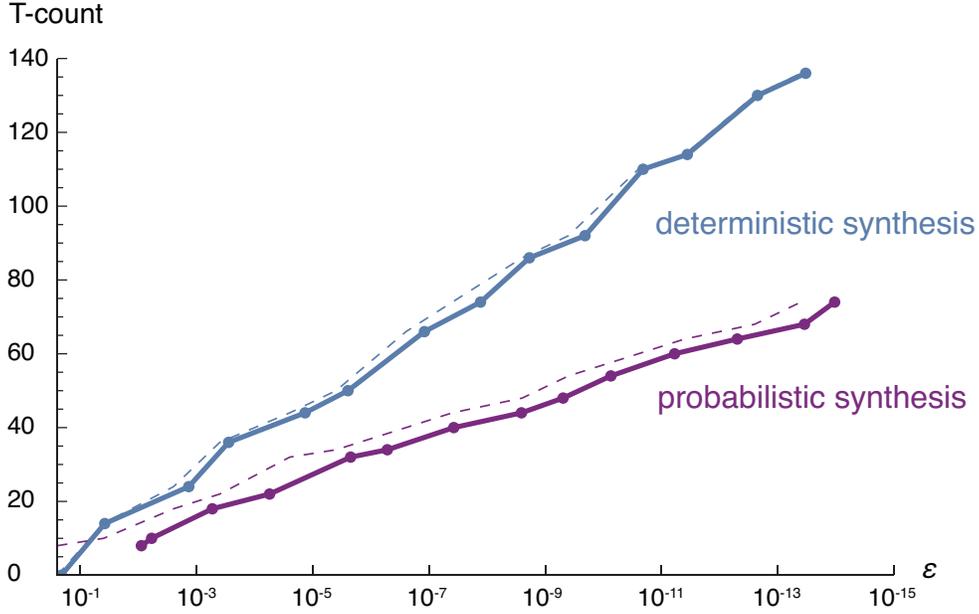}% Here is how to import EPS art
\caption{\label{fig:Tcount}  \textbf{Relationship between $T$-count and the approximation error for synthesizing $\ket{\phi}=\cos t\ket{0}+\sin t\ket{1}$ with $t=1$.} For each target approximation error, we run the Ross-Selinger algorithm to obtain a gate sequence to approximate $\phi$. The blue dashed line interpolates points, each of which represents a target approximation error and the $T$-count of the gate sequence. The actual approximation error and the $T$-count achieved by the gate sequence are plotted by blue dots. Note that both the target and actual approximation errors are represented by $\epsilon$. For each of the target approximation errors, we run the probabilistic synthesis algorithm and obtain a list of six gate sequences to be probabilistically sampled. The purple dashed line interpolates points, each of which represents a target approximation error and the maximum $T$-count of gate sequences in the list. The actual approximation error and the maximum $T$-count achieved by optimally mixing the gate sequence are plotted by purple dots.
}
\end{figure}

We assume that the set of elementary gates consists of Clifford gates and $T$-gate, which is a commonly utilized gate set in FTQC based on stabilizer codes or surface codes. Considering that the implementation cost of a $T$-gate is much higher than that of Clifford gates, it is necessary to minimize the $T$-count of the circuits. To do this, we use the Ross-Selinger algorithm \cite{RS16} to synthesize $R_y(2 t)=
\begin{pmatrix}
 \cos t&&-\sin t\\
 \sin t&& \cos t
\end{pmatrix}
$ and obtain a gate sequence that realizes a unitary operator $U_t(\simeq R_y(2 t))$. This allows us to obtain an approximated state $U_t\ket{0}(\simeq R_y(2 t)\ket{0}=\cos t\ket{0}+\sin t\ket{1})$
\footnote{Note that the target state can be generated by other unitary operators besides $R_y(2t)$: for example, $R_y(2 t)\ket{0}\propto R_y(2 t)R_z(t')\ket{0}$. Thus, we might be able to find a gate sequence with a lower $T$-count by synthesizing $R_y(2 t)R_z(t')$.
However, we are currently unaware of any algorithm that can synthesize $R_y(2 t)R_z(t')$ with an almost minimal $T$-count. In contrast, the Ross-Selinger algorithm can achieve an almost minimal $T$-count for synthesizing $R_y(2 t)$.}.
We run this deterministic synthesis algorithm for multiple randomly selected target states $\phi$ in $S_G$, with multiple target approximation errors.
By utilizing the description of each output gate sequence, we determine the $T$-count and the actual approximation error.

We perform probabilistic synthesis based on Theorem \ref{thm:main2} to synthesize the same multiple target states $\ket{\phi}$ with the same multiple target approximation errors.
When the target approximation error is $\epsilon$, we execute the Ross-Selinger algorithm within a target approximation error of $0.3\sqrt{\epsilon}$ for a $(0.7\sqrt{\epsilon})$-covering of $\{\psi\in S_G:\trdist{\psi-\phi}\leq2\sqrt{\epsilon}\}$. A set consisting of the target state $\ket{\phi_1}=\cos t\ket{0}+\sin t\ket{1}$ and two shifted states $\{\ket{\phi_{x}}\}_{x=2}^3=\{\cos t'\ket{0}+\sin t'\ket{1}:t'=t\pm 2\arcsin(0.7\sqrt{\epsilon})\}$ forms such a $(0.7\sqrt{\epsilon})$-covering when $\epsilon\leq0.07$. Thus, we obtain three gate sequences to generate states $\{\hat{\phi}_x\}_{x=1}^3$ after executing the Ross-Selinger algorithm.
To apply Theorem \ref{thm:main2}, we also require gate sequences to generate $\{\theta\hat{\phi}_x\theta\}_{x=1}^3$, the complex conjugation of $\{\hat{\phi}_x\}_{x=1}^3$. These gate sequences can be obtained by modifying the gate sequence to generate $\hat{\phi}_x$ without increasing the $T$-count. This is because $\theta T\theta\propto ZST$ and the set of Clifford gates is closed under the complex conjugation. After obtaining six synthesized states $\{\hat{\phi}_x,\theta\hat{\phi}_x\theta\}_{x=1}^3$, we solve the SDP described in Proposition \ref{prop:SDP} to determine the actual approximation error
\footnote{In principle, our SDP can determine the actual approximation error for an arbitrarily small target approximation error. However, due to the numerical error resulting from the SDP solver we use, we compute the actual approximation error based on the geometric interpretation shown in Fig.~\ref{fig:Bloch}, which outputs much more accurate solutions.}. Theorem \ref{thm:main2} guarantees the actual approximation error is smaller than $\epsilon$.
Note that without exploiting the symmetry of the target state, we need 13 states to form a $(0.7\sqrt{\epsilon})$-covering of $(2\sqrt{\epsilon})$-ball around $\phi$ due to the disk covering problem.

We examine how the $T$-count and the approximation error for a specific target state are related in Fig.~\ref{fig:Tcount}. 
As we can see, we were able to achieve a $50\sim60\%$ reduction in $T$-count. We observe similar behavior for other randomly selected target states (see \url{https://github.com/akibue/prob-synthesis} for details).

\subsection{Halving bit representation of pure states}
\label{sec:Cbit}
We verify that the existence of probabilistic and deterministic encoding given in Fig.~\ref{fig:Cencoding} can be reduced into a property of output states of the decoder $\Gamma$, as shown in the following propositions.
\begin{proposition}
\label{prop:probenc}
 A probabilistic encoding of $\puredop{\cd}$ with approximation error $\epsilon$ and a label set $X$ exists if and only if there exists set $\{\hat{\rho}_x\in\dop{\cd}\}_{x\in X}$ of mixed states satisfying
\begin{equation}
\label{eq:convcovering}
\max_{\phi\in\puredop{\cd}}\min_{p}\trdist{\phi-\sum_{x\in X}p(x)\hat{\rho}_x}\leq\epsilon,
\end{equation}
where the minimization is taken over a probability distribution $p$ over $X$.
\end{proposition}

\begin{proposition}
 A deterministic encoding of $\puredop{\cd}$ with approximation error $\epsilon$ and a label set $X$ exists if and only if there exists set $\{\hat{\rho}_x\in\dop{\cd}\}_{x\in X}$ of mixed states satisfying
\begin{equation}
\label{eq:covering}
\max_{\phi\in\puredop{\cd}}\min_{x\in X}\trdist{\phi-\hat{\rho}_x}\leq\epsilon.
\end{equation}
\end{proposition}
A set $\{\hat{\rho}_x\}_{x\in X}$ of mixed states satisfying Eq.~\eqref{eq:covering} is called an external $\epsilon$-covering of $\puredop{\cd}$. A set $\{\hat{\rho}_x\in\puredop{\cd}\}_{x\in X}$ of pure states satisfying Eq.~\eqref{eq:covering} is called an internal $\epsilon$-covering of $\puredop{\cd}$. The minimum size of internal (or external) $\epsilon$-coverings is called the internal (or external) covering number and denoted by $I_{\rm in}$ (or $I_{\rm ex}$). Note that $I_{\rm ex}\leq I_{\rm in}$ by definition and the minimum bit length $n_{\rm det}$ required for deterministic encodings is equal to $\lceil\log_2 I_{\rm ex}\rceil$.
We obtain the following lemma by using the volume consideration and applying the construction of an $\epsilon$-covering shown in \cite{ABmeetBanach}.
\begin{lemma}
 \label{lemma:detencoding}
For any $\epsilon\in\left(0,\frac{1}{2}\right]$ and an integer $d\geq2$ specified below,
the internal and external covering numbers $I_{\rm in}$ and $I_{\rm ex}$ of an $\epsilon$-covering of $\puredop{\cd}$ are bounded by 
\begin{equation}
2\cdot l(d,2\epsilon)\leq\log_2 I_{\rm ex}\leq \log_2I_{\rm in}\ \ \wedge\  \ 2\cdot l(d,\epsilon)\leq\log_2 I_{\rm in}\leq 2\cdot l(d,\epsilon)+\log_2(5d\ln d),
\end{equation}
where $l(d,\epsilon):=\left(d-1\right)\log_2\left(\frac{1}{\epsilon}\right)$. Moreover, if $d\geq4$, the first lower bound can be strengthened as $2\cdot l(d,\epsilon)\leq\log_2 I_{\rm ex}$.
\end{lemma}
The details of the proof are given in Appendix \ref{appendix:covering}. By combining this lemma with Theorem \ref{thm:main1}, we obtain the following theorem about the minimum bit length.
\begin{theorem}
\label{thm:main3}
For any $\epsilon\in\left(0,\frac{1}{2}\right]$ and an integer $d\geq2$ specified below,
the minimum bit length $n_{\rm det}$ (or $n_{\rm prob}$) of the deterministic (or probabilistic) encoding of $\puredop{\cd}$ with approximation error $\epsilon$ is bounded by 
\begin{eqnarray}
2\cdot l(d,2\epsilon)\leq &n_{\rm det}&\leq 2\cdot l(d,\epsilon)+\log_2(5d\ln d),\\
l(d,\epsilon)-\log_2d\leq&n_{\rm prob}&\leq l(d,\epsilon)+\log_2(5d\ln d),
\end{eqnarray}
where $l(d,\epsilon):=\left(d-1\right)\log_2\left(\frac{1}{\epsilon}\right)$. Moreover, if $d\geq4$, the first lower bound can be strengthened as $2\cdot l(d,\epsilon)\leq n_{\rm det}$.
\end{theorem}
\begin{proof}
 Since the bounds on $n_{\rm det}$ are a direct consequence of Lemma \ref{lemma:detencoding}, we show the bounds on $n_{\rm prob}$.
The upper bound is obtained by setting $\{\hat{\rho}_x\}_{x}$ in Proposition \ref{prop:probenc} to be the minimum internal $\sqrt{\epsilon}$-covering of $\puredop{\cd}$. This is because Theorem \ref{thm:main1} with $G=\{\idop\}$ guarantees that $\{\hat{\rho}_x\}_{x}$ satisfies Eq.~\eqref{eq:convcovering}, and an upper bound on the size of the internal $\sqrt{\epsilon}$-covering is given by Lemma \ref{lemma:detencoding}.
 
Next, we show the lower bound on $n_{\rm prob}$. Let $\{\hat{\rho}_x\in\dop{\cd}\}_{x\in X}$ satisfy Eq.~\eqref{eq:convcovering}. We obtain
\begin{eqnarray}
 \epsilon&\geq&\max_{\phi\in\puredop{\cd}}\min_p\trdist{\phi-\sum_{x\in X}p(x)\hat{\rho}_x}\geq\max_{\phi\in\puredop{\cd}}\min_p\left(1-\sum_xp(x)\tr{\phi\hat{\rho}_x}\right)\nonumber\\
 &=&1-\min_{\phi\in\puredop{\cd}}\max_{x\in X}\fidelity{\hat{\rho}_x}{\phi},
\end{eqnarray}
where we use $\trdist{\rho-\sigma}\geq\max_{\phi\in\puredop{\hh}}\tr{\phi(\rho-\sigma)}$ in the second inequality.

By letting $\hat{\rho}_x=\sum_{i=1}^{d}p(i|x)\phiI{i|x}$, we ensure that for any $\phi\in\puredop{\cd}$, there exists $i$ and $x$ such that
\begin{eqnarray}
 1-\epsilon\leq\fidelity{\hat{\rho}_x}{\phi}=\sum_{j=1}^{d}p(j|x)\fidelity{\phiI{j|x}}{\phi}\leq\fidelity{\phiI{i|x}}{\phi}=1-\trdist{\phi-\phiI{i|x}}^2.
\end{eqnarray}
Thus, $\{\phiI{i|x}\}_{i,x}$ is an internal $\sqrt{\epsilon}$-covering of $\puredop{\cd}$. Hence, the lower bound can be obtained by applying Lemma \ref{lemma:detencoding} as $\log_2(|X|d)\geq2\cdot l(d,\sqrt{\epsilon})=l(d,\epsilon)$.

\end{proof}

\if0
While this theorem implies the half reduction of a lower bound on the size of a synthesis circuit, the same reduction for $T$ gates is possible if we assume a conjecture given in \cite{RS16}. The conjecture states that the minimum number of $T$ gates for synthesizing arbitrary single qubit axial rotation is typically given as $3\log_2\left(\frac{1}{\epsilon}\right)+\Theta(1)$.
\fi

\subsection{Applications for analysis on entanglement measure}
\label{sec:application}
 Determining whether a quantum state $\rho$ is separable or entangled is a crucial inquiry in quantum information, as entanglement provides quantum advantages in various information processing tasks. The separability test is also fundamental to various optimization problems in distributed quantum computation. The separability test is computationally hard even if we are given the matrix representation of $\rho$ \cite{G10}. Further analysis of the computation complexity of the separability test has resulted in several important findings relating to QMA(2) \cite{BCY11-2, HM13, BKS17, HNW18}.
Although the separability test for general states is challenging, there are specific classes of states that make it easier to test for separability, e.g., low rank \cite{HHH96, LD11} and symmetric \cite{TAQLS18, CKMR07} states.

In order to identify the tractable states in the separability test, the study of the optimal convex approximation examines a generalized problem of how to approximate a target state $\rho$ with a probabilistic mixture of a restricted subset $\{\hat{\rho}_x\}_x$ of quantum states \cite{S17,LHYFH20,ZYY21conv}. When this subset consists of product states, it becomes the separability test.
From this general perspective, we demonstrated that restricting a target state to be rank-one or symmetry simplifies the optimization, as shown in Lemma \ref{lemma:mindist}.
Furthermore, we demonstrate that our general lemma for the optimal convex approximation can reproduce the nontrivial facts about entanglement, either already known or derivable through known facts, in a simpler and unified way.

Recall that the set of separable states is defined as follows.
\begin{definition}
 $\SEP:=\{\sigma\in\dop{\cd\otimes\cd}:\sigma=\sum_xp(x)\phi_x\otimes\psi_x\wedge \phi_x,\psi_x\in\puredop{\cd}\}$.
\end{definition}

In \cite{ANT22}, Girardin et al. used a neural network to conjecture Eqs.~\eqref{eq:entconj}.
Recall that $\rho^{\rm WER}_q\in\dop{\cd\otimes\cd}$ is the Werner state defined as $\rho_q^{\rm WER}:=\frac{2(1-q)}{d(d+1)}\Pi_\vee+\frac{2q}{d(d-1)}\Pi_\wedge$ with Hermitian projectors $\Pi_\vee$ and $\Pi_\wedge$ whose ranges are the symmetric subspace and antisymmetric subspace and $\rho^{\rm ISO}_q\in\dop{\cd\otimes\cd}$ is the isotropic state defined as $\rho_q^{\rm ISO}:=\frac{1-q}{d^2}\idop+q\Phi^+$ with $\ket{\Phi^+}=\frac{1}{\sqrt{d}}\sum_{i=0}^{d-1}\ket{ii}$, respectively. 
Since the Werner (or isotropic) state is entangled if and only if $\frac{1}{2}<q\leq1$ (or $\frac{1}{d+1}<q\leq1$), we assume they are entangled in Eqs.~\eqref{eq:entconj}.
 By exploiting the symmetry of the Werner (or isotropic) state and using Lemma \ref{lemma:mindist}, we can prove this conjecture. The complete proof is given in Appendix \ref{sec:entanglement}.

Note that Eqs.~\eqref{eq:entconj} can be proven straightforwardly by combining the following two facts: (i) the closest separable state can be assumed to be the Werner (or isotropic) state without loss of generality, and (ii) the Werner (or isotropic) state is separable if and only if $0\leq q\leq\frac{1}{2}$ (or $-\frac{1}{d^2-1}\leq q\leq\frac{1}{d+1}$).
In contrast, our proof directly computes the minimum trace distance without constructing the closest separable state, moreover, it includes a proof for (ii). Since a POVM element $M$ appeared in Eq.~\eqref{eq:witness1} can be regarded as an entanglement witness, our proof can be regarded as a method for ``quantifying entanglement with witness operators" \cite{BF05, BCY11}. Taking account of the fact that the closest separable state is not necessary in our method, it is expected that the advantage of our method becomes obvious when the closest separable state is unknown or analytically hard to obtain, as shown in the next example.

\if0
{\red We provide another exact formula for the distance from separable states by applying Lemma \ref{lemma:mindist} with another $G$ containing a conjugation.
\begin{proposition}
 For any probabilistic mixture $\rho\in\dop{\cdim{2}\otimes\cdim{2}}$ of maximally entangled states, it holds that
 \begin{equation}
 \min_{\sigma\in \SEP}\trdist{\rho-\sigma}=\frac{1}{2}C(\rho),
\end{equation}
where $C(\rho)$ is the concurrence of $\rho$.
\end{proposition}
\begin{proof}
 It is known that $\rho$ is can be transformed into a Bell-diagonal state
 \begin{equation}
 \rho'=\sum_{i=1}^4p_i\Psi_i
\end{equation}
with $p_1\geq p_2\geq p_3\geq p_4$ by local unitary transformations. It has been shown that $C(\rho)=C(\rho')=\max\{0,p_1-p_2-p_3-p_4\}$ \cite{W98}. We can verify $\rho'$ and the set of pure product states are invariant under the action of $G$ generated by $\{\Theta,\sigma_X\otimes\sigma_X,\sigma_Z\otimes\sigma_Z\}$. In this case, $P_G:=\{P\in\pos{\hh}:\forall U\in G,[U,P]=0\}=\{r_1\idop\otimes\idop+r_2\sigma_X\otimes\sigma_X+r_3\sigma_Y\otimes\sigma_Y+r_4\sigma_Z\otimes\sigma_Z:r_i\in\rr\}$. By using Eq.~\eqref{eq:witness1}, we obtain
\begin{eqnarray}
  \min_{\sigma\in \SEP}\trdist{\rho'-\sigma}=
\end{eqnarray}
\end{proof}
}
\fi

Due to its clear operational meaning, the resource measure based on trace distance has been investigated for various resource theories, including entanglement and coherence \cite{BP14}. Lemma \ref{lemma:mindist} provides an alternate concise proof for the following recently identified coincidence between entanglement and coherence measures.
\begin{proposition} \cite[Theorem 3]{JSNCS16}
\label{prop:coherenceentanglement}
For pure states $\ket{\Phi}=\sum_{i=0}^{d-1}\alpha_i\ket{ii}$ and $\ket{\phi}=\sum_{i=0}^{d-1}\alpha_i\ket{i}$, it holds that
\begin{equation}
\label{eq:coherenceentanglement}
 \min_{\sigma\in \SEP}\trdist{\Phi-\sigma}=\min_{\rho\in I}\trdist{\phi-\rho},
\end{equation}
where $I:=\conv{\{\ketbra{i}\}_{i=0}^{d-1}}$ is called a set of incoherent states and $\{\ket{i}\}_{i=0}^{d-1}$ is an orthonormal basis.
\end{proposition}

Since it is suggested that a simple closed-form formula for Eq.~\eqref{eq:coherenceentanglement} might not exist \cite{JSNCS16}, the closest separable state is also hard to obtain. However, our method is applicable to show the relationship of the minimum approximation error between different types of probabilistic approximation  by exploiting the purity of the target states. Moreover, it simplifies the proof of \cite[Theorem 3]{JSNCS16}. The complete proof is given in Appendix \ref{sec:entanglement}.

\section{Discussion}
\label{sec:conclusion}
We investigated the limitation of the optimal probabilistic state synthesis and its potential for reducing the size of a synthesis circuit.
As a main result, we verified the tight relationship between the approximation error obtained by the optimal probabilistic state synthesis and the optimal deterministic one.
We also constructed an efficient method to convert a deterministic synthesis algorithm into a probabilistic one that quadratically reduces the approximation error.

To estimate how the error reduction reduces the size of a synthesis circuit, we performed a numerical simulation and evaluated the length of the classical bit string required to approximately encode a pure state. As a result, we found that probabilistic encoding asymptotically halves the bit length. Note that under the presence of noise on elementary gates, which was not taken into account in this study, certain conditions on the noise may be required to achieve the quadratic reduction of the approximation error.
However, our SDP can still be used to numerically determine the optimal probabilistic synthesis in cases where the noise is explicitly described.

In addition to our contribution to the state synthesis, the our result would improve the performance of classical simulation of a quantum computer as well as that of optimization algorithms including a brute force search over pure states, e.g., the separability test \cite{ITCE04}. This is because we essentially show that the set of pure states can be approximated by its $\epsilon$-covering or probabilistic mixtures of its $\sqrt{\epsilon}$-covering in the same accuracy, where the size of the minimum $\sqrt{\epsilon}$-covering is almost the square root of that of the minimum $\epsilon$-covering.

These results are based on general theorems about the optimal convex approximation of a quantum state. While the optimal convex approximation and state synthesis have been studied in different contexts, our theorems have demonstrated that analyzing the former problem provides not only the fundamental limitation of probabilistic synthesis but also a construction of an efficient synthesis algorithm. Furthermore, our theorems contribute to the original motivation of the studies of the optimal convex approximation \cite{S17,LHYFH20,ZYY21conv}, which is quantifying a resource measure in convex resource theories \cite{HO13, BG15, CG19} such as the resource theory of entanglement. Indeed, the SDP constructed in Proposition \ref{prop:SDP} would provide a basis for numerical investigation for such resource measures. Our theorems would reveal more quantitative relationships between different resource measures as shown in Proposition \ref{prop:coherenceentanglement}.

\section*{Data availability}
Numerical results together with instructions on how to reproduce them, are available online at \url{https://github.com/akibue/prob-synthesis}.

\section*{Code availability}
 The code is available from the corresponding author on request.

\bibliography{references_npj.bib}% Produces the bibliography via BibTeX.

\begin{acknowledgments}
We thank Masato Koashi, Yuki Takeuchi, Yasunari Suzuki, Yasuhiro Takahashi, and Adel Sohbi for their helpful discussions.
This work was partially supported by JST Moonshot R\&D MILLENNIA Program (Grant no.JPMJMS2061).
SA was partially supported by JST, PRESTO Grant no.JPMJPR2111 and MEXT Q-LEAP Grant no. JPMXS0120319794.
GK was supported in part by the Grant-in-Aid for Scientific Research (C) no.20K03779, (C) no.21K03388, and (S) no.18H05237 of JSPS, and CREST (Japan Science and Technology Agency) Grant no.JPMJCR1671.
ST was partially supported by JSPS KAKENHI Grant nos. JP20H05966 and JP22H00522.
\end{acknowledgments}

\section*{Author contributions}
The ideas were given by S.A. All the authors contributed to the preparation of the manuscript and verification of proofs.

%\section*{Competing interests}
%The authors declare no competing interests.

\newpage
\appendix
\section{Formal SDP}
\label{appendix:SDP}
After briefly restating the formal definition of semidefinite programming \cite{WBook}, we show the formal SDPs provided in Proposition \ref{prop:SDP} and verify their strong duality.
\subsection{Preliminaries}
Let $\Xi:\linop{\hh_1}\rightarrow\linop{\hh_2}$ be a linear Hermitian-preserving mapping and $A$ and $B$ be Hermitian operators on $\hh_1$ and $\hh_2$, respectively.
SDP is an optimization problem formally defined with a triple $(\Xi,A,B)$ as follows:
\begin{equation}
\begin{tabular}{rlcrl}
\multicolumn{2}{c}{\underline{{\rm Primal problem}}} &\ \ \ \ \ \ \ \ \ \ \ \ \ \  
 &\multicolumn{2}{c}{\underline{{\rm Dual problem}}}\\
{\rm maximize:}&$\tr{AX}$&&{\rm minimize:}&$\tr{BY}$ \\
{\rm subject to:}& $X\in\pos{\hh_1}$,&&
{\rm subject to:}& $Y\text{\ is\ a\ Hermitian\ operator\ on\ }\hh_2$,\\
&$\Xi(X)=B$&&&$\Xi^\dag(Y)\geq A$,
\end{tabular} 
\end{equation}
where $\Xi^\dag:\linop{\hh_2}\rightarrow\linop{\hh_1}$ is the adjoint of $\Xi$, defined as the linear mapping satisfying $\tr{Y^\dag\Xi(X)}=\tr{(\Xi^\dag(Y))^\dag X}$ for all $X\in\linop{\hh_1}$ and $Y\in\linop{\hh_2}$.
We can easily verify that the solution to the primal problem is smaller than or equal to that of the dual problem. The situation when the two solutions coincide is called a strong duality. Slater's theorem states that the strong duality holds if either of the following conditions are met:
\begin{enumerate}
 \item The solution to the primal problem is finite, and there exists a Hermitian operator $Y$ on $\hh_2$ such that $\Xi^\dag(Y)>A$.
 \item The solution to the dual problem is finite, and there exists a positive definite operator $X$ on $\hh_1$ such that $\Xi(X)=B$.
\end{enumerate}

\subsection{Formal SDP and verification of strong duality}
A formal SDP to compute $\trdist{\rho-\sigma}$ is defined with a triple $(\Xi,A,B)$ such that
\begin{eqnarray}
A=\left(
\begin{matrix}
 \rho-\sigma&0\\
 0&0
\end{matrix}
\right),\ \ 
B=\idop,\ \ 
\Xi\left(\left(
\begin{matrix}
 M&*\\
 *&M'
\end{matrix}
\right)\right)=
M+M'
\end{eqnarray}
holds for any linear operators $M,M'\in\linop{\hh}$, where the asterisks in the argument to $\Xi$ represent arbitrary linear operators upon which $\Xi$ does not depend, and we identify a linear operator and its matrix representation with respect to a fixed orthonormal basis. The dual problem is obtained by observing that the adjoint of $\Xi$ satisfies
\begin{equation}
 \Xi^\dag\left(Y\right)=
\left(
\begin{matrix}
 Y&0\\
 0&Y
 \end{matrix}
\right)
\end{equation}
for any linear operator $Y\in\linop{\hh}$. We can verify the strong duality of this SDP by observing $\Xi\left(\frac{\idop}{2}\oplus\frac{\idop}{2}\right)=B$ and applying Slater's theorem.

The formal SDP shown in Proposition \ref{prop:SDP} is defined with a triple $(\Xi,A,B)$ such that
\begin{eqnarray}
A&=&\left(
\begin{matrix}
 \rho&0&0&0\\
 0&0&0&0\\
 0&0&0&0\\
 0&0&0&-1
\end{matrix}
\right)\\
B&=&\left(
\begin{matrix}
 \idop&0\\
 0&0
\end{matrix}
\right)\\
  \Xi^\dag\left(\left(
\begin{matrix}
 Y&*\\
*&P
\end{matrix}
\right)\right)&=&
\left(
\begin{matrix}
 Y+\sum_{x\in X}P(x)\hat{\rho}_x&0&0&0\\
 0&Y&0&0\\
 0&0&P&0\\
 0&0&0&-\tr{P}
\end{matrix}
\right)
\end{eqnarray}
holds for any linear operators $Y\in\linop{\hh}$ and $P\in\linop{\cdim{|X|}}$,
where $P(x)$ represents a diagonal element $\bra{x}P\ket{x}$. The primal problem is obtained by observing that the adjoint of $\Xi^\dag$ satisfies
\begin{equation}
 \Xi\left(
 \left(
\begin{matrix}
 M&*&*&*\\
 *&M'&*&*\\
 *&*&Q&*\\
 *&*&*&t
\end{matrix}
\right)
 \right)=
 \left(
\begin{matrix}
 M+M'&0\\
 0&\sum_{x\in X}\tr{M\hat{\rho}_x}\ketbra{x}+Q-t\idop
\end{matrix}
\right)
\end{equation}
for any linear operators $M,M'\in\linop{\hh}$ and $Q\in\linop{\cdim{|X|}}$.
We can verify the strong duality of this SDP by observing $\Xi\left(\frac{\idop}{2}\oplus\frac{\idop}{2}\oplus\frac{\idop}{2}\oplus 1\right)=B$ and applying Slater's theorem.

\section{Proof of Lemma \ref{lemma:support}}
\label{sec:proof_support}
\begin{proof}
We show the nontrivial inequality $(L.H.S.)\geq(R.H.S.)$. We also consider the nontrivial case when $\epsilon<\frac{1}{2}$. By using Lemma \ref{lemma:mindist}, we obtain
 \begin{eqnarray}
 \label{eq:supp1}
 (L.H.S.)&=&\max_{\psi\in S_G}\left(\fidelity{\psi}{\phi}-\max_{x\in X}\fidelity{\psi}{\hat{\phi}_x}\right)=\max_{\psi\in S_G}\left(\min_{x\in X}\trdist{\psi-\hat{\phi}_x}^2-\trdist{\psi-\phi}^2\right),\\
 \label{eq:supp2}
  (R.H.S.)&=&\max_{\psi\in S_G}\left(\fidelity{\psi}{\phi}-\max_{x\in \hat{X}}\fidelity{\psi}{\hat{\phi}_x}\right)=\max_{\psi\in S_G}\left(\min_{x\in \hat{X}}\trdist{\psi-\hat{\phi}_x}^2-\trdist{\psi-\phi}^2\right),
\end{eqnarray}
where we use the right equality in Ineq.~\eqref{ineq:FG} in the last equality.

If $\psi$ in $\eball{\epsilon}{\phi}:=\{\psi\in S_G:\trdist{\psi-\phi}\leq \epsilon\}$ maximizes Eq.~\eqref{eq:supp2}, there exists $x\in \hat{X}$ such that $\min_{x\in X}\trdist{\psi-\hat{\phi}_x}=\trdist{\psi-\hat{\phi}_x}$ since $\min_{x\in X}\trdist{\psi-\hat{\phi}_x}\leq\epsilon$  and $\trdist{\phi-\psi}\leq\epsilon$. This implies $Eq.~\eqref{eq:supp1}\geq Eq.~\eqref{eq:supp2}$.

If $\psi\in S_G\setminus\eball{\epsilon}{\phi}$ maximizes Eq.~\eqref{eq:supp2}, we can show $(R.H.S.)=0$ as follows.
First, if there exists a unitary operator $W\in G$ such that $W\ket{\phi}=\omega\ket{\phi}$,  $W\ket{\psi}=\omega'\ket{\psi}$ and $\omega\neq\omega'$, $\trdist{\phi-\psi}=1$ since $\ket{\phi}$ and $\ket{\psi}$ are eigenvectors associated with distinct eigenvalues. Thus, $(R.H.S.)=0$.
Next, we assume $W\ket{\phi}=\omega\ket{\phi}$ and  $W\ket{\psi}=\omega\ket{\psi}$ for any unitary operator $W\in G$ with some $\omega\in\cc$.
If there exist an antiunitary operator $V\in G$, we fix global phases of $\ket{\phi}$ and $\ket{\psi}$ to satisfy the following three conditions: (i) $V\ket{\phi}=\ket{\phi}$, (ii) $V\ket{\psi}=\ket{\psi}$ and (iii) $\braket{\phi}{\psi}\geq0$. For there exist $\ket{\phi'}$ and $\ket{\psi'}$ satisfying $V\ket{\phi'}=\ket{\phi'}$ and $V\ket{\psi'}=\ket{\psi'}$, $\braket{\phi'}{\psi'}=\bra{\phi'}V\ket{\psi'}=\bra{\psi'}V^\dag\ket{\phi'}=\braket{\psi'}{\phi'}$ implies $\braket{\phi'}{\psi'}\in\rr$, and we can choose $\ket{\phi}$ (or $\ket{\psi}$) from $\{\pm\ket{\phi'}\}$ (or $\{\pm\ket{\psi'}\}$) to satisfy the three conditions (i)-(iii). 
In this case, we can verify that $V'\ket{\phi}=\omega'\ket{\phi}$ and $V'\ket{\psi}=\omega'\ket{\psi}$ for any antiunitary operator $V'\in G$ with some $\omega'\in\cc$ since $\ket{\phi}$ and $\ket{\psi}$ are eigenvectors of a unitary operator $V'V\in G$ associated with the same eigenvalue. 
If $G$ contains no antiunitary operator, we can set arbitrary global phases for $\ket{\phi}$ and $\ket{\psi}$ such that $\braket{\phi}{\psi}\geq0$ holds.
Then, we can verify that any pure state $\ket{\eta}=a\ket{\phi}+b\ket{\psi}$ with real coefficients $a,b\in\rr$ satisfies $\eta\in S_G$ since $U\ket{\eta}=\omega\ket{\eta}$ for any unitary or antiunitary operator $U\in G$ with some $\omega\in\cc$.

Let $\ket{\psi}=\cos t_1\ket{\phi}+\sin t_1\ket{\phi_\bot}$, where $\braket{\phi}{\phi_\bot}=0$ and $t_1\in\left[0,\frac{\pi}{2}\right]$. 
Note that $\phi_\bot\in S_G$ and $\psi\notin\eball{\epsilon}{\phi}$ implies $\sin t_1>\epsilon$. 
Define $\ket{\hat{\phi}}=\cos t_2\ket{\phi}+\sin t_2\ket{\phi_\bot}$ and $\ket{\hat{\phi}_\bot}=-\sin t_2\ket{\phi}+\cos t_2\ket{\phi_\bot}$, where $\sin t_2=\epsilon$ and $t_2\in\left[0,\frac{\pi}{2}\right]$.
Since $\hat{\phi}\in S_G$ and $\{\hat{\phi}_x\}_{x}$ is an $\epsilon$-covering of $S_G$, there exists $x\in X$ such that $\trdist{\hat{\phi}-\hat{\phi}_x}\leq\epsilon$.
Moreover, $x\in\hat{X}$ since $\trdist{\hat{\phi}-\phi}=\epsilon$. 
By letting $(\hat{\phi}+\hat{\phi}_\bot)\ket{\hat{\phi}_x}=a\ket{\hat{\phi}}+\beta\ket{\hat{\phi}_\bot}$ with $a\geq\sqrt{1-\epsilon^2}$ and $\beta\in\cc$, we obtain
\begin{eqnarray}
 \min_{x\in \hat{X}}\trdist{\psi-\hat{\phi}_x}^2&\leq&\trdist{\psi-\hat{\phi}_x}^2=1-\fidelity{\psi}{\hat{\phi}_x}\\
 &=&1-\left|a\cos(t_1-t_2)+\beta\sin(t_1-t_2)\right|^2\\
 &\leq&1-\re{a\cos(t_1-t_2)+\beta\sin(t_1-t_2)}^2\\
 \label{eq:supportproof1}
 &=&1-\left(
\begin{pmatrix}
 a\\\re{\beta}
\end{pmatrix}\cdot
\begin{pmatrix}
 \cos(t_1-t_2)\\\sin(t_1-t_2)
\end{pmatrix}
\right)^2,
\end{eqnarray}
where $\re{\beta}$ represents the real part of a complex number $\beta$. We can draw the possible region of $
\begin{pmatrix}
 a\\\re{\beta}
\end{pmatrix}
$ as the shaded region in Fig.~\ref{fig:support}. By using $0< t_1-t_2\leq\frac{\pi}{2}-t_2$ and Fig.~\ref{fig:support}, we obtain
\begin{equation}
 Eq.~\eqref{eq:supportproof1}\leq1-\left(
\begin{pmatrix}
 \cos t_2\\-\sin t_2
\end{pmatrix}\cdot
\begin{pmatrix}
 \cos(t_1-t_2)\\\sin(t_1-t_2)
\end{pmatrix}
\right)^2=1-\fidelity{\phi}{\psi}=\trdist{\psi-\phi}^2.
\end{equation}
This implies $(R.H.S.)=0$, which completes the proof.

\begin{figure}[h]
\includegraphics[height=.2\textheight]{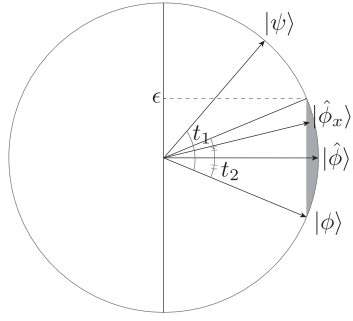}% Here is how to import EPS art
\caption{\label{fig:support} \textbf{Relationship between $\ket{\psi}$, $\ket{\phi}$, $\ket{\hat{\phi}}$ and $\ket{\hat{\phi}_x}$.} $\ket{\psi}$, $\ket{\phi}$, $\ket{\hat{\phi}}$ and $\ket{\hat{\phi}_x}$ correspond to $
\begin{pmatrix}
 \cos(t_1-t_2)\\\sin(t_1-t_2)
\end{pmatrix}
$, $
\begin{pmatrix}
 \cos t_2\\-\sin t_2
\end{pmatrix}
$, $
\begin{pmatrix}
 1\\0
\end{pmatrix}
$ and $
\begin{pmatrix}
 a\\\re{\beta}
\end{pmatrix}
$, respectively.}
\end{figure}

\end{proof}

\section{Probabilistic state synthesis for a high dimensional system}
\label{sec:high dimension}
 Note that our method assumes a target state is taken from a constant-dimensional Hilbert space, similar to the Solovay-Kitaev algorithm. 
We discuss the difficulty and the possibility of applying our method to high-dimensional systems.
Our method becomes impractical when attempting to synthesize a general target state $\phi$ in $\cd$ with a large dimension $d$. This is because
the size $|\tilde{X}|$ of a $c\epsilon$-covering $\{\phi_x\}_{x\in\tilde{X}}$ of a $2\epsilon$-ball around $\phi$ ($c<1$), used in our method, grows exponentially with respect to $d$.
However, in some applications such as machine learning, $\phi$ may have a sparse structure, i.e., it resides in a small-dimensional subspace.
If $\phi$ is taken from a constant-dimensional subspace $\vv$, then it is symmetric under the action of $G=\{\idop,2\Pi-\idop\}$ with Hermitian projector $\Pi$ whose range is $\vv$. In this situation, $|\tilde{X}|$ remains constant since a $2\epsilon$-ball around $\phi$ consists of pure states in $\vv$ (assuming $\epsilon<\frac{1}{2}$). By utilizing an efficient deterministic state synthesis algorithm proposed for $\phi$ with the sparse structure \cite{GH21, XTX22} and assuming that $U=2\Pi-\idop$ is efficiently implementable, we obtain gate sequences to generate $\{\hat{\rho}_x\}_{x\in\hat{X}}$, used in the proof of Theorem \ref{thm:main2}. To compute the optimal probability distribution $p(x)$ sampling $\{\hat{\rho}_x\}_{x\in\hat{X}}$, we need to solve the SDP proposed in Proposition \ref{prop:SDP}. Since each $\hat{\rho}_x$ can be replaced by $\Pi\hat{\rho}_x\Pi$ due to Lemma \ref{lemma:mindist}, the size of the SDP also remains constant. Computing the matrix representation of $\Pi\hat{\rho}_x\Pi$ from the gate sequence to generate $\hat{\rho}_x$ is generally difficult, but, if we achieve this, we can perform a probabilistic state synthesis that enables the quadratic error reduction.

\section{Proof of Lemma \ref{lemma:detencoding}}
\label{appendix:covering}
\begin{itemize}
 \item {\bf Proof for} $2(d-1)\log_2\left(\frac{1}{\epsilon}\right)\leq\log_2 I_{\rm in}$: Let $\mu$ be the unitarily invariant probability measure on the Borel sets of $\puredop{\cd}$. As shown in subsection \ref{appendix:vol1}, the volume of $\epsilon$-ball $\eball{\epsilon}{\phi}:=\left\{\psi\in\puredop{\cd}:\trdist{\psi-\phi}\leq\epsilon\right\}$ can be calculated as $\mu(\eball{\epsilon}{\phi})=\epsilon^{2(d-1)}$ for any $\phi\in\puredop{\cd}$. Since the union of $\eball{\epsilon}{\hat{\phi}_x}$ covers $\puredop{\cd}$ when $\{\hat{\phi}_x\}_{x\in X}$ is an $\epsilon$-covering of $\puredop{\cd}$, we obtain $\mu(\eball{\epsilon}{\phi})|X|\geq1$.
 
 \item {\bf Proof for} $2(d-1)\log_2\left(\frac{1}{\epsilon}\right)\leq\log_2 I_{\rm ex}$ if $d\geq4$ and $2(d-1)\log_2\left(\frac{1}{2\epsilon}\right)\leq\log_2 I_{\rm ex}$ if $d\in\{2,3\}$: Similar to the previous proof, this bound is a consequence of the upper bound on the volume of $\epsilon$-ball $\eball{\epsilon}{\rho}:=\left\{\psi\in\puredop{\cd}:\trdist{\psi-\rho}\leq\epsilon\right\}$ shown in subsection \ref{appendix:vol2}.

 \item {\bf Proof for} $\log_2 I_{\rm in}\leq 2(d-1)\log_2\left(\frac{1}{\epsilon}\right)+\log_2(5d\ln d)$:
 We construct an internal $\epsilon$-covering ($\epsilon\in(0,1]$) following the proof in \cite[Corollary 5.5]{ABmeetBanach}. The construction is based on the fact that sufficiently many pure states randomly sampled form an $\epsilon$-covering. However, since the probability of a new random pure state residing in the uncovered region decreases when many random $\epsilon$-balls are sampled, it is better to stop sampling a pure state and change the construction strategy.

In the proof, we represent some parameters explicitly, which are tailored to the $\epsilon$-covering with respect to the trace distance.
 Assume $d\geq2$ and let $D=2(d-1)(\geq2)$. Let $\{\phiI{j}\in\puredop{\cd}\}_{j=1}^{J_R}$ be a set of finite randomly sampled pure states with respect to product measure $\mu^{J_R}$. The expected volume of the region not covered by $A:=\cup_{j=1}^{J_R}\eball{\epsilon_R}{\phiI{j}}$ ($0<\epsilon_R\leq 1$) can be calculated as follows:
 \begin{eqnarray}
 &&\int d\mu^{J_R}\mu\left(A^c\right)\nonumber\\
 &&= \int d\mu^{J_R}\int d\mu(\psi)\prod_{j=1}^{J_R}\indicator{\trdist{\psi-\phiI{j}}>\epsilon_R}\nonumber\\
 &&=\int d\mu(\psi)\prod_{j=1}^{J_R} \int d\mu(\phiI{j})\indicator{\trdist{\psi-\phiI{j}}>\epsilon_R}\nonumber\\
 &&=\left(1-\epsilon_R^D\right)^{J_R}\leq\exp\left(-J_R\epsilon_R^D\right),
\end{eqnarray}
where we use Fubini's theorem and Eq.~\eqref{eq:volume_of_eball} in the second and the third equations, respectively. Note that $\indicator{X}\in\{0,1\}$ is the indicator function, i.e., $\indicator{X}=1$ iff $X$ is true.

Thus, there exists $\{\phiI{j}\}_{j=1}^{J_R}$ such that $\mu\left(A^c\right)\leq\exp\left(-J_R\epsilon_R^D\right)$. Pick $\{\psiI{j}\}_{j=1}^{J_P}$ as much as possible such that $\eball{\epsilon_P}{\psiI{j}}$ are disjoint and contained in $A^c$. When $0<\epsilon_P\leq\epsilon_R\leq 1$, we can verify that $\{\phiI{j}\}_{j=1}^{J_R}\cup\{\psiI{j}\}_{j=1}^{J_P}$ is an $(\epsilon_R+\epsilon_P)$-covering  and its size $J:=J_R+J_P$ is upper bounded as
\begin{equation}
 J\leq J_R+\frac{\exp\left(-J_R\epsilon_R^D\right)}{\epsilon_P^D}.
\end{equation}
By setting $J_R=\left\lceil\frac{D}{\epsilon_R^D}\ln\left(\frac{\epsilon_R}{\epsilon_P}\right)\right\rceil$, $\epsilon_P=\frac{\epsilon_R}{x}$, and $\epsilon_R=\frac{x}{1+x}\epsilon$ with $x\geq1$, we obtain the following upper bound:
\begin{equation}
 J\leq\left\lceil \frac{D\ln x}{\epsilon_R^D}\right\rceil+\frac{1}{\epsilon_R^D}
\leq\frac{1}{\epsilon^D}\left\{\left(1+\frac{1}{x}\right)^D(D\ln x+1)+1\right\}=\frac{2d\ln d}{\epsilon^D}\cdot\frac{\alpha(d,x)}{2d\ln d},
\end{equation}
where $\alpha(d,x)=\left(1+\frac{1}{x}\right)^D(D\ln x+1)+1$.
Since $\lim_{d\rightarrow\infty}\frac{\alpha(d,D\ln D)}{2d\ln d}=1$, we obtain that for any $r>2$ there exists $d_0\in\nn$ such that
\begin{equation}
\forall d\geq d_0,\forall\epsilon\in(0,1], J\leq rd(\ln d)\left(\frac{1}{\epsilon}\right)^{2(d-1)}.
\end{equation}
For example, if $r=5$, we can set $d_0=2$.
This completes the proof. 
 
\end{itemize}

\subsection{Volume analysis 1}
\label{appendix:vol1}
We compute the volume of $\eball{\epsilon}{\phi}$ in $\puredop{\cd}$ as follows:
\begin{equation}
\label{eq:volume_of_eball}
\forall d\in\nn,\forall \epsilon\in(0,1],\forall \phi\in\puredop{\cd}, \mu(\eball{\epsilon}{\phi})=\epsilon^{2(d-1)},
\end{equation}
where $\mu$ is the unitarily invariant probability measure on the Borel sets of $\puredop{\cd}$.

When $d=1$, Eq.~\eqref{eq:volume_of_eball} holds. By assuming $d\geq2$, we proceed as follows:
\begin{eqnarray}
&&\mu(\eball{\epsilon}{\phi})\nonumber\\
&&=\mu\left(\left\{\psi\in\puredop{\cd}:\trdist{\ketbra{0}-\psi}\leq\epsilon\right\}\right)\nonumber\\
&&=\mu\left(\left\{\psi\in\puredop{\cd}:|\braket{0}{\psi}|^2\geq1-\epsilon^2\right\}\right)\nonumber\\
&&=\xi\left(\left\{\vec{x}\in\rdim{2d}:\lpnorm{2}{\vec{x}}=1\wedge x_1^2+x_2^2\geq1-\epsilon^2\right\}\right),
\end{eqnarray}
where the first equality uses fixed pure state $\ket{0}$ and the unitary invariance of $\mu$ and the trace distance, the second equality uses Eq.~\eqref{ineq:FG}, and the third equality uses the relationship between $\mu$ and the uniform spherical probability measure $\xi$. Using a spherical coordinate system, we can proceed as follows:
\begin{eqnarray}
\mu(\eball{\epsilon}{\phi})&=&\frac{V(\epsilon)}{V(1)},\\
\text{where}\ V(\epsilon)&:=&\int_{D_\epsilon}\sin^{2d-2}\theta\sin^{2d-3}\phi d\theta d\phi\nonumber\\
&=&4\int_{\hat{D}_\epsilon}\sin^{2d-2}\theta\sin^{2d-3}\phi d\theta d\phi
\end{eqnarray}
and the domain of the integration $D_\epsilon$ is given by $\{(\theta,\phi):\theta,\phi\in(0,\pi),\sin\theta\sin\phi<\epsilon\}$. Since this domain and that of the integrand have reflection symmetries for two lines $\theta=\frac{\pi}{2}$ and $\phi=\frac{\pi}{2}$, it is sufficient to perform the integration in domain $\hat{D}_\epsilon:=\{(\theta,\phi):\theta,\phi\in\left(0,\frac{\pi}{2}\right),\sin\theta\sin\phi<\epsilon\}$. By changing the variables as
$ \begin{pmatrix}
 x\\y
\end{pmatrix}
=
\begin{pmatrix}
 \sin\theta\sin\phi\\
 \sin\theta
\end{pmatrix}$, we obtain
\begin{eqnarray}
 V(\epsilon)&=&4\int_0^{\epsilon}dxx^{2d-3}\int_x^1dy\frac{y}{\sqrt{1-y^2}\sqrt{y^2-x^2}}\nonumber\\
 &=&4\int_0^{\epsilon}dxx^{2d-3}\left[\arcsin\sqrt{\frac{1-y^2}{1-x^2}}\right]_1^x\nonumber\\
 &=&\frac{\pi}{d-1}\epsilon^{2(d-1)}
\end{eqnarray}
for $\epsilon\in[0,1]$. This completes the calculation.

\subsection{Volume analysis 2}
\label{appendix:vol2}
We show the following upper bound on the volume of $\epsilon$-ball $\eball{\epsilon}{\rho}$. For any $\epsilon\in\left(0,\frac{1}{2}\right]$, it holds that
\begin{eqnarray}
\label{appp:volbound1}
  \mu(\eball{\epsilon}{\rho})&\leq&  (2\epsilon)^{2(d-1)}\ \ \text{for}\  d\in\{1,2,3\}\\
  \label{appp:volbound2}
    \mu(\eball{\epsilon}{\rho})&\leq & \epsilon^{2(d-1)}\ \ \ \ \ \text{for}\ d\geq4.
\end{eqnarray}
This bound and $\mu(\eball{\epsilon}{\phi})=\epsilon^{2(d-1)}$ imply that the volume of the $\epsilon$-ball can be maximized by setting its center as a pure state if $d\geq4$, which is contrary to what happens in a qubit ($d=2$), where $\eball{\epsilon}{\rho}$ corresponds to the intersection of the Bloch sphere and a ball centered at $\rho$ and the intersection is maximized not by a ball centered at a point on the Bloch sphere but by a ball centered at a point inside the Bloch ball. The qubit case also implies that the condition $d\geq4$ for the second inequality cannot be fully relaxed. $\mu(\eball{\epsilon}{\sigma})=1$ if $\epsilon\geq1-\frac{1}{d}$ with the maximally mixed state $\sigma=\frac{1}{d}\idop$ implies that another condition $\epsilon\in\left(0,\frac{1}{2}\right]$ is also not fully removable. 

{\bf Proof of Eq.~\eqref{appp:volbound1}}: 
By defining $\phi:=\arg\min_{\phi\in\puredop{\cd}}\trdist{\phi-\rho}$, we obtain $\eball{\epsilon}{\rho}\subseteq\eball{2\epsilon}{\phi}$. This is because $\trdist{\psi-\phi}\leq\trdist{\psi-\rho}+\trdist{\phi-\rho}\leq2\trdist{\psi-\rho}<2\epsilon$ for any pure state $\psi\in\eball{\epsilon}{\rho}$. This completes the proof because $\mu(\eball{\epsilon}{\rho})\leq\mu(\eball{2\epsilon}{\phi})=(2\epsilon)^{2(d-1)}$ for any $\epsilon\in\left(0,\frac{1}{2}\right]$.

{\bf Proof of Eq.~\eqref{appp:volbound2}}: 
We assume $d\geq4$.
Let $\rho=\sum_{i=0}^{d-1}p_i\ketbra{i}$, where $\{\ket{i}\}_i$ is a set of eigenvectors of $\rho$ and eigenvalues are arranged in decreasing order, i.e., $p_0\geq p_1\geq \cdots$. Since $\mu(\eball{\epsilon}{\rho})$ depends not on the eigenvectors but on the eigenvalues of $\rho$, it is sufficient to consider only diagonal $\rho$ with respect to a fixed basis. However, it is difficult to precisely calculate $\mu(\eball{\epsilon}{\rho})$ due to a complicated relationship between $\psi$ and the largest eigenvalue of $\psi-\rho$, resulting from the condition $\epsilon\geq\trdist{\psi-\rho}=\lambda_{\max}(\psi-\rho)$. 

We derive the lower bound $f_\rho(\psi)$ of $\trdist{\psi-\rho}$ and use the relationship $\mu(\eball{\epsilon}{\rho})\leq\mu\left(\{\psi:f_\rho(\psi)\leq\epsilon\}\right)$ to show Eq.~\eqref{appp:volbound2}, where $f_\rho$ is a measurable function. Since the simple bound $f_\rho(\psi)=1-\fidelity{\psi}{\rho}$ is too loose to show Eq.~\eqref{appp:volbound2}, we derive a tighter lower bound as follows: 
Let $\Pi$ and $\Pi^\bot$ be the Hermitian projectors on two-dimensional subspace $\vv\supseteq\vspan{\{\ket{0},\ket{\psi}\}}$ and its orthogonal complement, respectively. We then obtain
\begin{eqnarray}
\trdist{\psi-\rho}&\geq&\trdist{\Pi(\psi-\rho)\Pi+\Pi_\bot(\psi-\rho)\Pi_\bot}\nonumber\\
\label{eq:fpsi}
&=&\trdist{\psi-\Pi\rho\Pi}+\trdist{\Pi_\bot\rho\Pi_\bot},
\end{eqnarray}
where we use the monotonicity of the trace distance under a CPTP mapping in the first inequality. Define $f_\rho(\psi)$ as the value in Eq.~\eqref{eq:fpsi}, which can be explicitly written as
 \begin{equation}
 f_\rho(\psi)= \frac{1}{2}\sqrt{(1+p_0-q)^2-4(p_0-q)|\braket{0}{\psi}|^2}+\frac{1}{2}(1-p_0-q),
\end{equation}
where $q=\bra{0_\bot}\rho\ket{0_\bot}$ and $\{\ket{0},\ket{0_\bot}\}$ is an orthonormal basis of $\vv$. The explicit formula implies $f_\rho$ is  uniquely defined (although neither $\vv$ nor $q$ is uniquely defined if $\psi=\ketbra{0}$) and continuous, and thus measurable. 

\begin{figure}[h]
\includegraphics[height=.23\textheight]{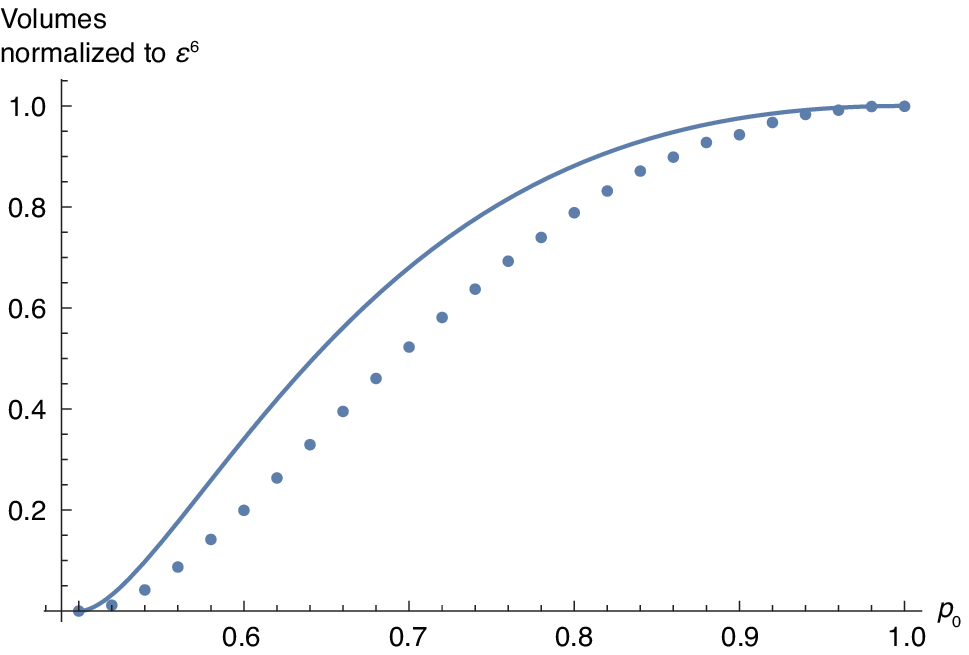}% Here is how to import EPS art
\caption{\label{fig:boundcomparison} \textbf{Plots of estimated values of $\mu(\eball{\frac{1}{2}}{\rho})$ (dots) and $g_{4,\frac{1}{2}}(p_0)$ (curve) for $\rho=p_0\ketbra{0}+(1-p_0)\ketbra{1}\in\dop{\cdim{4}}$.} $\mu(\eball{\frac{1}{2}}{\rho})$ is estimated by uniformly sampling $10^7$ pure states. The plots indicate $\mu(\eball{\frac{1}{2}}{\rho})$ is accurately upper bounded by $g_{4,\frac{1}{2}}(p_0)$.}
\end{figure}

We assume $\epsilon\in\left(0,\frac{1}{2}\right]$.
Since $\mu(\eball{\epsilon}{\rho})=0$, satisfying Eq.~\eqref{appp:volbound2}, if $p_0\leq1-\epsilon$, we consider the case $p_0>1-\epsilon$. In this case, we obtain
\begin{equation}
 f_\rho(\psi)\leq \epsilon\Leftrightarrow
 |\braket{0}{\psi}|^2\geq \frac{(\epsilon+p_0)(1-q-\epsilon)}{p_0-q},
\end{equation}
where the condition is not trivial, i.e., $\frac{(\epsilon+p_0)(1-q-\epsilon)}{p_0-q}\in(0,1)$.
We calculate an upper bound on $\mu(\eball{\epsilon}{\rho})$ as follows: 
By defining $\unitary{\hh}$ as the set of unitary operators on $\hh$ and $C:\unitary{\cdim{d-1}}\rightarrow\unitary{\cd}$ as $C(U):=\ketbra{0}\oplus U$, we can show that for any unitary operator $U\in\unitary{\cdim{d-1}}$,
\begin{eqnarray}
 &&\mu\left(\{\psi\in\puredop{\cd}:f_\rho(\psi)\leq\epsilon\}\right)\nonumber\\
&&=\mu\left(\{\psi:f_\rho(C(U)\psi C(U)^\dag)\leq\epsilon\}\right)\nonumber\\
\label{eq:uinv}
&&= \int_{\puredop{\cd}}d\mu(\psi)\indicator{|\braket{0}{\psi}|^2\geq\frac{(\epsilon+p_0)(1-q_U-\epsilon)}{p_0-q_U}},
\end{eqnarray}
where we use the unitary invariance of $\mu$ in the first equality, $q_U=\bra{0_\bot} C(U)^\dag\rho C(U)\ket{0_\bot}$, and $\indicator{X}\in\{0,1\}$ is the indicator function, i.e., $\indicator{X}=1$ iff $X$ is true. By integrating Eq.~\eqref{eq:uinv} with respect to the unitarily invariant probability measure on $\unitary{\cdim{d-1}}$ and using Fubini's theorem, we obtain
 \begin{equation}
 \mu\left(\{\psi\in\puredop{\cd}:f_\rho(\psi)<\epsilon\}\right)=\int_{\puredop{\cd}}d\mu(\psi)\int_{\puredop{\cdim{d-1}}}d\mu(\phi) \indicator{|\braket{0}{\psi}|^2\geq\frac{(\epsilon+p_0)(1-\fidelity{\rho}{\phi}-\epsilon)}{p_0-\fidelity{\rho}{\phi}}},
\end{equation}
where $\phi\in\puredop{\cdim{d-1}}$ is identified with a pure state on $\puredop{\cd}$ acting on subspace $\vspan{\{\ket{1},\cdots,\ket{d-1}\}}$. Using Fubini's theorem again and Eq.~\eqref{eq:volume_of_eball}, we can proceed with the calculation:
\begin{eqnarray}
 &&=\int_{\puredop{\cdim{d-1}}}d\mu(\phi)\delta(\fidelity{\rho}{\phi})^{2(d-1)}\nonumber\\
 &&=\int_{\puredop{\cdim{d-1}}}d\mu(\phi)\delta\left(\sum_{i=1}^{d-1}p_i|\braket{i}{\phi}|^2\right)^{2(d-1)}\nonumber\\
 &&\leq\int_{\puredop{\cdim{d-1}}}d\mu(\phi)\delta\left((1-p_0)|\braket{1}{\phi}|^2\right)^{2(d-1)}\nonumber\\
 &&=(d-2)\int_0^1(1-x)^{d-3}\delta((1-p_0)x)^{2(d-1)}dx=:g_{d,\epsilon}(p_0),
\end{eqnarray}
where $\delta(q)=\sqrt{\frac{(\epsilon+q)(p_0+\epsilon-1)}{p_0-q}}$. We use the convexity of $\delta^{2(d-1)}$ and the unitary invariance of $\mu$ in the last inequality, and we use the probability density of $x=|\braket{1}{\phi}|^2$ derived by Eq.~\eqref{eq:volume_of_eball} in the last equality.
To confirm the calculation, we plot a comparison between $\mu(\eball{\epsilon}{\rho})$ and its upper bound $g_{d,\epsilon}(p_0)$ for a particular $\rho$ in Fig.~\ref{fig:boundcomparison}, where we use the following explicit expression of $g_{4,\epsilon}(p_0)$:
\begin{eqnarray}
 \label{eq:g4expression}
 g_{4,\epsilon}(p_0)&=&2(p_0+\epsilon-1)^3\bigg\{\frac{1-6b-ab^2}{2ab^2}+\frac{3(a+1)}{a(b-a)}\left(1-\frac{a}{b-a}\log\frac{b}{a}\right)\bigg\},
\end{eqnarray}
where $a=\frac{2p_0-1}{\epsilon+p_0}$, $b=\frac{p_0}{\epsilon+p_0}$, and $p_0\in(1-\epsilon,1)$. Note that $g_{4,\epsilon}(1)=\epsilon^6=\lim_{p_0\rightarrow1}g_{4,\epsilon}(p_0)$.

It is sufficient to show that under the two conditions $\epsilon\in\left(0,\frac{1}{2}\right]$ and $d\geq4$,
\begin{equation}
\forall p_0\in(1-\epsilon,1),\frac{dg_{d,\epsilon}}{dp_0}\geq0,
\end{equation}
since $g_{d,\epsilon}(1)=\epsilon^{2(d-1)}$. Since the integrand of $g_{d,\epsilon}$ and its partial derivative with respect to $p_0$ are continuous, we can interchange the partial differential and integral operators:
\begin{eqnarray}
 \frac{dg_{d,\epsilon}}{dp_0}&=&(d-2)\int_0^1(1-x)^{d-3}\frac{\partial}{\partial p_0}\delta((1-p_0)x)^{2(d-1)}dx\nonumber\\
 &=&\alpha_{d,\epsilon}(p_0)\int_0^1\beta_{\epsilon}(p_0,x)\gamma_{d,\epsilon}(p_0,x)dx,
\end{eqnarray}
where $\alpha_{d,\epsilon}(p_0)=(d-2)(d-1)(p_0+\epsilon-1)^{d-2}$, $\beta_{\epsilon}(p_0,x)=-(1-p_0)^2x^2+(1-\epsilon-\epsilon^2-p_0^2)x+(1-\epsilon)\epsilon$ and $\gamma_{d,\epsilon}(p_0,x)=\frac{(1-x)^{d-3}(\epsilon+(1-p_0)x)^{d-2}}{(p_0-(1-p_0)x)^d}$. Since $\alpha_{d,\epsilon}$ and $\gamma_{d,\epsilon}$ are non-negative in the entire considered region $R:=\{(p_0,x):p_0\in(1-\epsilon,1)\wedge x\in[0,1]\}$, $\frac{dg_{d,\epsilon}}{dp_0}\geq0$ if $\beta_{\epsilon}$ is non-negative for all $x\in[0,1]$. However, $\beta_{\epsilon}$ can be negative for some $x\in[0,1]$ if and only if $\beta_{\epsilon}(p_0,1)<0$. Taking account of considered region $R$, it is sufficient to show $\frac{dg_{d,\epsilon}}{dp_0}\geq0$ for all $p_0\in\left(\frac{1+\sqrt{1-4\epsilon^2}}{2},1\right)(\subseteq(1-\epsilon,1))$, where $\beta_{\epsilon}$ can be negative. 

For fixed $p^*\in\left(\frac{1+\sqrt{1-4\epsilon^2}}{2},1\right)$, let $x^*\in(0,1)$ satisfy $\beta_{\epsilon}(p^*,x^*)=0$. Since $\beta_{\epsilon}(p^*,x)$ is monotonically decreasing in $x\geq0$, $x^*$ is uniquely defined, $\beta_{\epsilon}(p^*,x)>0$ if $x\in[0,x^*)$ and $\beta_{\epsilon}(p^*,x)<0$ if $x\in(x^*,1]$. Thus, showing
\begin{equation}
\label{eq:condition1}
 \forall d\geq4,\exists c>0,
 \left\{
\begin{array}{ll}
 \gamma_{d+1,\epsilon}(p^*,x)\geq c\gamma_{d,\epsilon}(p^*,x)&\text{for}\ x\in[0,x^*)\\
  \gamma_{d+1,\epsilon}(p^*,x)\leq c\gamma_{d,\epsilon}(p^*,x)&\text{for}\ x\in(x^*,1]
\end{array}
 \right.
\end{equation}
and
\begin{equation}
\label{eq:condition2}
\frac{dg_{4,\epsilon}}{dp_0}\bigg|_{p_0=p^*}\geq0
\end{equation}
is sufficient for $\forall d\geq4,\frac{dg_{d,\epsilon}}{dp_0}\Big|_{p_0=p^*}\geq0$.
This is because
\begin{eqnarray}
 \alpha_{d+1,\epsilon}(p^*)^{-1}\frac{dg_{d+1,\epsilon}}{dp_0}\bigg|_{p_0=p^*}&=&\int_0^{x^*}\beta_{\epsilon}(p^*,x)\gamma_{d+1,\epsilon}(p^*,x)dx+\int_{x^*}^{1}\beta_{\epsilon}(p^*,x)\gamma_{d+1,\epsilon}(p^*,x)dx\nonumber\\
 &\geq& c\biggl\{\int_0^{x^*}\beta_{\epsilon}(p^*,x)\gamma_{d,\epsilon}(p^*,x)dx+\int_{x^*}^{1}\beta_{\epsilon}(p^*,x)\gamma_{d,\epsilon}(p^*,x)dx\biggr\}\nonumber\\
 &=&c\alpha_{d,\epsilon}(p^*)^{-1}\frac{dg_{d,\epsilon}}{dp_0}\bigg|_{p_0=p^*}
\end{eqnarray}
holds for any $d\geq4.$

First, we show Eq.~\eqref{eq:condition1}. By observing that for any $d\geq4$,
\begin{equation}
\gamma_{d+1,\epsilon}(p^*,x)-c \gamma_{d,\epsilon}(p^*,x)= \gamma_{d,\epsilon}(p^*,x)\left(\frac{(1-x)(\epsilon+(1-p^*)x)}{p^*-(1-p^*)x}-c\right),
\end{equation}
 $h_{\epsilon,p^*}(x):=\frac{(1-x)(\epsilon+(1-p^*)x)}{p^*-(1-p^*)x}$ is monotonically decreasing in $x\in[\hat{x},1]$ and $h_{\epsilon,p^*}(x)\geq h_{\epsilon,p^*}(\hat{x})$ for $x\in[0,\hat{x}]$ with $\hat{x}:=\max\left\{0,1-\frac{2p^*-1}{p^*(1-p^*)}\epsilon\right\}(\leq x^*)$, setting $c=h_{\epsilon,p^*}(x^*)(>0)$ implies Eq.~\eqref{eq:condition1}.

Next, Eq.~\eqref{eq:condition2} can be verified by using the explicit expression in Eq.~\eqref{eq:g4expression}.

\section{Proof of entanglement measures}
\label{sec:entanglement}
\begin{proof}[Proof of  Eqs.~\eqref{eq:entconj}]
 The Werner (or isotropic) state is the only state that is invariant under the action of $u\otimes u$ (or $u\otimes u^*$), where $u$ is a unitary operator on $\cd$. Thus, by setting $G=\{u\otimes u\}$ (or $G=\{u\otimes u^*\}$), we obtain $\rho^{\rm WER}_q\in P_G=\{a\Pi_\vee+b\Pi_\wedge:a,b\geq0\}$ (or $\rho^{\rm ISO}_q\in P_G=\{a\idop+(b-a)\Phi^+:a,b\geq0\}$), where $P_G:=\{P\in\pos{\hh}:\forall U\in G,[U,P]=0\}$.
By setting $\{\hat{\rho}_x\}_{x}=\{\phi\otimes\psi:\phi,\psi\in\puredop{\cd}\}$, which is invariant under the action of $G$, we can verify that the minimum approximation error in Eq.~\eqref{eq:witness1} coincides with $ \min_{\sigma\in \SEP}\trdist{\rho-\sigma}$.
Rigorously speaking, we cannot directly apply Lemma \ref{lemma:mindist} since neither $G$ nor $\{\hat{\rho}_x\}_{x}$ are finite sets, but their compactness allows us to extend Lemma \ref{lemma:mindist} and apply it. Details of such extension are provided in Appendix \ref{appendix:compactextension}.

For the Werner state, we obtain
\begin{eqnarray}
  \min_{\sigma\in \SEP}\trdist{\rho_q^{\rm WER}-\sigma}&=&\max_{0\leq a,b\leq 1}\left(\tr{(a\Pi_\vee+b\Pi_\wedge)\rho_q^{\rm WER}}-\max_{\phi,\psi\in\puredop{\cd}}\tr{(a\Pi_\vee+b\Pi_\wedge)(\phi\otimes\psi)}\right)\nonumber\\\\
  &=&\max_{0\leq a,b\leq 1}\left(a(1-q)+bq-\frac{a+b}{2}-\max_{\phi,\psi\in\puredop{\cd}}\frac{a-b}{2}\fidelity{\phi}{\psi}\right)\\
  &=&
  \left\{
\begin{array}{ll}
 q-\frac{1}{2}&{\rm if\ }\frac{1}{2}<q\leq1\\
 0&{\rm if\ }0\leq q\leq\frac{1}{2}
\end{array}\right.
\end{eqnarray}
where the last equality is obtained by setting $(a,b)=(0,1)$ or $a=b$ when $q>\frac{1}{2}$ or $q\leq\frac{1}{2}$, respectively.

For the isotropic state, we obtain
\begin{eqnarray}
  \min_{\sigma\in \SEP}\trdist{\rho_q^{\rm ISO}-\sigma}&=&\max_{0\leq a,b\leq 1}\left(\tr{(a\idop+(b-a)\Phi^+)\rho_q^{\rm ISO}}-\max_{\phi,\psi\in\puredop{\cd}}\tr{(a\idop+(b-a)\Phi^+)(\phi\otimes\psi)}\right)\nonumber\\\\
  &=&\max_{0\leq a,b\leq 1}\left(a+(b-a)\left(\frac{1-q}{d^2}+q\right)-a-\max_{\phi,\psi\in\puredop{\cd}}\frac{b-a}{d}\fidelity{\phi}{\psi}\right)\\
 &=&
   \left\{
\begin{array}{ll}
 \frac{d^2-1}{d^2}\left(q-\frac{1}{d+1}\right)&{\rm if\ }\frac{1}{d+1}<q\leq1\\
 0&{\rm if\ }-\frac{1}{d^2-1}\leq q\leq\frac{1}{d+1}
\end{array}\right.
\end{eqnarray}
where the last equality is obtained by setting $(a,b)=(0,1)$ or $a=b$ when $q>\frac{1}{d+1}$ or $q\leq\frac{1}{d+1}$, respectively.
\end{proof}

\begin{proof}[Proof of Proposition \ref{prop:coherenceentanglement}]
By using the extended Lemma \ref{lemma:mindist} (see Appendix \ref{appendix:compactextension}) with $G=\{u\otimes u^*,U_{SWAP}(u\otimes u^*)\}$ and $\{\hat{\rho}_x\}_{x}=\{\phi\otimes\psi:\phi,\psi\in\puredop{\cd}\}$, where $U_{SWAP}$ is the swap operator, $u$ is taken from a set $\left\{\sum_{k=0}^{d-1}e^{i\theta_k}\ketbra{k}:\forall k,\theta_k\in\{1,e^{i\frac{2\pi}{3}},e^{i\frac{4\pi}{3}}\}\right\}$ of diagonal unitary operators and $u^*$ is the complex conjugate of $u$ with respect to the basis $\{\ket{k}\}_{k=0}^{d-1}$,
\begin{eqnarray}
 (L.H.S.)&=&\max_{\ket{\Psi}\in\vspan{\{\ket{ii}\}_i}}\left(\tr{\Psi\Phi}-\max_{\phi,\psi\in\puredop{\cd}}\tr{\Psi(\phi\otimes\psi)}\right)\nonumber\\
 &=&\max_p\left(\left(\sum_{i=0}^{d-1}\sqrt{p(i)}|\alpha_i|\right)^2-\max_ip(i)\right),
\end{eqnarray}
where we obtain the last equality by setting $\ket{\Psi}=\sum_{i=0}^{d-1}\sqrt{p(i)}\frac{\alpha_i}{|\alpha_i|}\ket{ii}$.

By using Eq.~\eqref{eq:witness2} with $G=\{\idop\}$ and $\{\hat{\rho}_x\}_{x}=\{\ketbra{i}\}_{i=0}^{d-1}$, we obtain
\begin{eqnarray}
 (R.H.S.)=\max_{\psi\in\puredop{\cd}}\left(\tr{\psi\phi}-\max_{i}\bra{i}\psi\ket{i}\right)=\max_p\left(\left(\sum_{i=0}^{d-1}\sqrt{p(i)}|\alpha_i|\right)^2-\max_ip(i)\right),
\end{eqnarray}
where we set $\ket{\psi}=\sum_{i=0}^{d-1}\sqrt{p(i)}\frac{\alpha_i}{|\alpha_i|}\ket{i}$ for the maximization.
This completes the proof.
\end{proof}

\section{Extension of Lemma \ref{lemma:mindist}}
\label{appendix:compactextension}
In this section, we extend Lemma \ref{lemma:mindist} for the case when both $G$ and $X$ are infinite sets. In the following extended lemma, we assume the Borel set $\mathcal{B}_G$ is defined in a subgroup $G$ of unitary and antiunitary operators on $\hh$. Since the Borel set in $G$ is the smallest $\sigma$-algebra containing the topology $\tau$ in $G$, we define $\tau$ as follows. First, we can verify that the set $\vv$ of summations of linear and antilinear operators is a complex vector space. We define a norm in the space as follows: $\lpnorm{}{A+B}:=\max_{\phi\in\puredop{\hh},t\in\rr}\lpnorm{2}{(A+B)e^{it}\ket{\phi}}$, where $A$ and $B$ are a linear and antilinear operator, respectively. Since $\lpnorm{}{A+B}^2=\max_{\phi\in\puredop{\hh}}\left(\lpnorm{2}{A\ket{\phi}}^2+\lpnorm{2}{B\ket{\phi}}^2+2|\bra{\phi}A^\dag B\ket{\phi}|\right)$, we can verify that a nontrivial condition of a norm, $\lpnorm{}{A+B}=0\Leftrightarrow A=B=0$.
Since $G$ is a subset of a normed linear space $\vv$, we can define a topology $\tau$ in $G$ induced by the norm $\lpnorm{}{\cdot}$.
Note that a function $f:G\rightarrow\linop{\hh}$ defined as $f(U):= U^\dag AU$ with $A\in\linop{\hh}$ is continuous.

{\bf Lemma \ref{lemma:mindist}.} (extended version)
{\it
 Let $G$ be a subgroup of unitary and antiunitary operators where a $G$-invariant probability measure $\mu:\mathcal{B}_G\rightarrow[0,1]$ is defined, i.e., $\mu(UE)=\mu(EU)=\mu(E)$ for any $U\in G$ and Borel set $E\in\mathcal{B}_G$.
Let $P_G$ be the set of positive semidefinite operators invariant under the action of $G$, i.e., $P_G:=\{P\in\pos{\hh}:\forall U\in G,[U,P]=0\}$.
  If $\rho\in P_G\cap\dop{\hh}$ and a compact set $\{\hat{\rho}_x\in\dop{\hh}\}_{x\in X}$ of mixed states is invariant under the action of $G$, i.e., $\{\hat{\rho}_x\}_{x\in X}=\{U\hat{\rho}_x U^\dag\}_{x\in X}$ for all $U\in G$, it holds that
\begin{equation}
\label{eq:exwitness1}
 \min_{\sigma\in\conv{\{\hat{\rho}_x\}_{x}}}\trdist{\rho-\sigma}=\max_{\substack{0\leq M\leq\idop\\ M\in P_G}}\left(\tr{M\rho}-\max_{x\in X}\tr{M\hat{\rho}_x}\right),
\end{equation}
where $\conv{A}$ represents the set of finite convex combinations of a set $A$.
In particular, when $\rho$ is a pure state $\phi$, it holds that
 \begin{equation}
 \label{eq:exwitness2}
 \min_{\sigma\in\conv{\{\hat{\rho}_x\}_{x}}}\trdist{\phi-\sigma}=\max_{\psi\in P_G\cap\puredop{\hh}}\left(\tr{\psi\phi}-\max_{x\in X}\tr{\psi\hat{\rho}_x}\right).
\end{equation}
}

\begin{proof}
We start from a mixed state $\rho$. Since we can apply the minimax theorem due to the compactness of $\conv{\{\hat{\rho}_x\}_{x}}$, we obtain
\begin{equation}
\label{eq:exsubspacelemma1}
 (L.H.S.\  of\  Eq.~\eqref{eq:exwitness1})=\max_{0\leq M\leq\idop}\left(\tr{M\rho}-\max_{x\in X}\tr{M\hat{\rho}_x}\right).
\end{equation}
This proves $(L.H.S.)\geq(R.H.S.)$.
Let $M$ maximize Eq.~\eqref{eq:exsubspacelemma1}. Due to the invariance of $\rho$ and $\{\hat{\rho}_x\}_x$ under the action of $G$, we can verify that $U^\dag MU$ also maximizes Eq.~\eqref{eq:exsubspacelemma1}. By defining $\hat{M}=\int_GU^\dag MU d\mu(U)$, we obtain
\begin{eqnarray}
 (R.H.S.\  of\  Eq.~\eqref{eq:exwitness1})&\geq&\tr{\hat{M}\rho}-\max_{x\in X}\tr{\hat{M}\hat{\rho}_x}
=\tr{M\rho}-\max_{x\in X}\left(\int_{G}\tr{MU\hat{\rho}_xU^\dag}d\mu(U)\right)\nonumber\\\\
&\geq&\tr{M\rho}-\int_G\max_{x\in X}\tr{MU\hat{\rho}_xU^\dag}d\mu(U)
=\tr{M\rho}-\max_{x\in X}\tr{M\hat{\rho}_x}\\
&=&(L.H.S.\  of\  Eq.~\eqref{eq:exwitness1}).
\end{eqnarray}

We can prove the case when $\rho$ is a pure state $\phi$ in the same way as the proof of Lemma \ref{lemma:mindist}.

\end{proof}

Note that $\mu$ is the counting measure when $G$ is finite. Thus, when $X$ is also finite, this extended lemma is reduced into the original Lemma \ref{lemma:mindist}.
Let $\unitary{\cd}$ and $\nu$ be the set of unitary operators on $\cd$ and the unitarily invariant probability measure on $\unitary{\cd}$, respectively.
In Section \ref{sec:application}, we use this extended lemma in the following three cases:
\begin{enumerate}
 \item When $G=\{u\otimes u: u\in\unitary{\cd}\}$, we can verify that $\mu(E):=\nu\left(\{u\in\unitary{\cd}:u\otimes u\in E\}\right)$ is a $G$-invariant probability measure.
 
  \item When $G=\{u\otimes u^*: u\in\unitary{\cd}\}$, we can verify that $\mu(E):=\nu\left(\{u\in\unitary{\cd}:u\otimes u^*\in E\}\right)$  is a $G$-invariant probability measure.

\end{enumerate}

\end{document}